\documentclass[%
reprint,
superscriptaddress,
twocolumn
 amsmath,amssymb,
 aip,
 jcp,
 citeautoscript,
]{revtex4-2}

\usepackage[utf8]{inputenc}
\usepackage[T1]{fontenc}
\usepackage{graphicx} 
\usepackage{float}
\usepackage{array}
\usepackage{charter}
\usepackage[usenames,dvipsnames]{xcolor}
\usepackage{setspace}
\usepackage{hyperref}

\begin{document}

\title{Long-range dispersion-inclusive machine learning potentials for structure search and optimization of hybrid organic-inorganic interfaces}% 

\author{Julia Westermayr}
\affiliation{Department of Chemistry, University of Warwick, Coventry, CV4 7AL, United Kingdom}
\author{Shayantan Chaudhuri}
\affiliation{Department of Chemistry, University of Warwick, Coventry, CV4 7AL, United Kingdom}
\altaffiliation{Centre for Doctoral Training in Diamond Science and Technology, University of Warwick, Coventry, CV4 7AL, United Kingdom}
\author{Andreas Jeindl}
\affiliation{Institute of Solid State Physics, Graz University of Technology, 8010 Graz, Austria}
\author{Oliver T. Hofmann}
\affiliation{Institute of Solid State Physics, Graz University of Technology, 8010 Graz, Austria}
\author{Reinhard J. Maurer}
\affiliation{Department of Chemistry, University of Warwick, Coventry, CV4 7AL, United Kingdom}
\email{r.maurer@warwick.ac.uk}

\begin{abstract}
The computational prediction of the structure and stability of hybrid organic-inorganic interfaces provides important insights into the measurable properties of electronic thin film devices, coatings, and catalyst surfaces and plays an important role in their rational design. However, the rich diversity of molecular configurations and the important role of long-range interactions in such systems make it difficult to use machine learning (ML) potentials to facilitate structure exploration that otherwise require computationally expensive electronic structure calculations. We present an ML approach that enables fast, yet accurate, structure optimizations by combining two different types of deep neural networks trained on high-level electronic structure data. The first model is a short-ranged interatomic ML potential trained on local energies and forces, while the second is an ML model of effective atomic volumes derived from atoms-in-molecules partitioning. The latter can be used to connect short-range potentials to well-established density-dependent long-range dispersion correction methods. For two systems, specifically gold nanoclusters on diamond (110) surfaces and organic $\pi$-conjugated molecules on silver (111) surfaces, we train models on sparse structure relaxation data from density functional theory and show the ability of the models to deliver highly efficient structure optimizations and semi-quantitative energy predictions of adsorption structures.
\end{abstract}

\maketitle

\section{Introduction}
Surface nanostructures play a fundamental role in medicine,~\cite{Dong2021BP,Gewin2021N} solar cell and fuel cell technologies,~\cite{Guo2016S,Cano2018NE} and photo- or electrocatalysis.~\cite{Li2021MH,Bottari2017JEC} Several strategies exist to form nanostructures, such as DNA-directed assembly,~\cite{Lalander2010ACSN} electrodeposition,~\cite{Bottari2017JEC} or self-assembly at hybrid organic-inorganic interfaces.\cite{Lloyd2016NL} The molecular composition and molecule-surface interaction strength crucially determine the surface structures that are formed ~\cite{Jeindl2021ACSN,Otero2011AM,Tan2019JPCM} and the nucleation and initial growth of nanoclusters (NCs) are crucial steps in controlling a nanostructures' final morphology,~\cite{Bottari2017JEC,Cobb2018ARAC} which itself is important for tuning catalytic selectivity and activity.~\cite{Kelly2003JPCB} A better understanding of surface nanostructures can thus advance a wide variety of research fields.~\cite{He2019Ns,Hofmann2021PCCP}

%problem
Electronic structure theory plays a vital role in the characterization and exploration of organic-inorganic interfaces and materials, but is limited by intrinsic errors such as the lack of long-range dispersion interactions in common density functionals~\cite{Grimme2016CR,Hermann2017CR,Maurer2016PSS} and the high computational effort associated with the intrinsic length scale of surface structures. The former issue has been addressed in recent years with the emergence of efficient and accurate long-range dispersion correction methods such as the Grimme and Tkatchenko-Scheffler (TS) families of methods.~\cite{Tkatchenko2009PRL,Grimme2016CR} In the case of metal-organic interfaces, the vdW$^{\mathrm{surf}}$~\cite{Ruiz2012PRL} and many-body dispersion (MBD)\cite{Tkatchenko2012PRL,Ambrosetti2014JCP} methods, in combination with generalized gradient approximations (GGAs) or range-separated hybrid functionals, have been shown to provide highly accurate predictions of adsorption structures and stabilities.~\cite{Maurer2015JCP,Maurer2016PSS,benzenesymm, bloweyAlkaliDopingLeads2020,Hoermann2020JCP,Vydrov2006JCP,Karolewski2013JCP,Tan2019JPCM,Otero2011AM} Reliable identification and optimization of structures at metal-organic interfaces is a particular challenge due to the structural complexity and the large number of degrees of freedom (molecular orientation, adsorption site, coverage),~\cite{Hofmann2021PCCP} which creates a particular need for structural exploration methods that are efficient. Examples of simulation methods that can alleviate computational effort compared to DFT include semi-empirical electronic structure methods, such as density functional tight-binding (DFTB),~\cite{Hourahine2020JCP} which usually provides a good compromise between accuracy and computational efficiency. Recently, DFTB has been coupled with the vdW and MBD methods~\cite{Stoehr2016JCP, Hourahine2020JCP} to incorporate long-range dispersion, but unfortunately few reliable DFTB parametrizations for metal-organic interfaces exist to date.~\cite{Fihey2015JCC}

%ML
Machine learning-based interatomic potentials (MLIPs) offer high computational efficiency whilst retaining the accuracy of the underlying training data based on electronic structure theory. Atomistic MLIP methods include Gaussian Approximation Potentials~\cite{Bartok2010PRL,Bartok2015,Young2021CS} or neural network (NN) potentials (e.g. SchNet,~\cite{Schutt2018,Schutt2017_double,Schuett2019JCTC} PhysNet~\cite{Unke2019JCTC} or Behler-Parinello type NNs~\cite{Behler2015IJQC,Behler2017ACIE,Behler2007}), which describe atoms in their chemical and structural environment within a cutoff region. MLIPs have the potential to advance structure searches,~\cite{Jorgensen2019,PhysRevB.102.075427,alphafold} geometry optimizations,~\cite{Meyer2020,Yang2021JCP} and molecular dynamics (MD) simulations~\cite{Behler2017ACIE,Chmiela2018NC,Bogojeski2020NC,Deringer2021N} of highly complex and large-scale systems comprising many thousands of atoms.\cite{Kulik2022ES} However, most established MLIP approaches learn short-range interactions between atoms by introducing a radial cutoff within which the atomic interactions are captured. This can lead to challenges when attempting to capture long-range electrostatic or dispersion interactions.~\cite{Unke2019JCTC}  Recent attempts of accounting for long-range interactions in MLIPs have explicitly treated them as separate additive contributions to the potential,~\cite{Morawietz2013JPCA,Unke2019JCTC,Kun2018CS,Unke2021NC} such as the third and higher generation NN potentials of Behler and co-workers,~\cite{Ko2021ACr,Ko2021NC} where a charge-equilibration scheme was introduced. These approaches have been demonstrated to accurately describe MD or spectroscopic signatures,~\cite{Kun2018CS} small clusters on surfaces~\cite{Ko2021NC}, water dimers\cite{zhang2021deep} and clusters~\cite{Morawietz2013JPCA}, crystals,~\cite{zhang2021deep} and phase diagrams.~\cite{muhli2021machine} However, they are often limited to single systems and lack a transferable description of potential energy surfaces, especially long-range interactions.  

\begin{figure}[ht]
    \centering
    \includegraphics[width=3.2in]{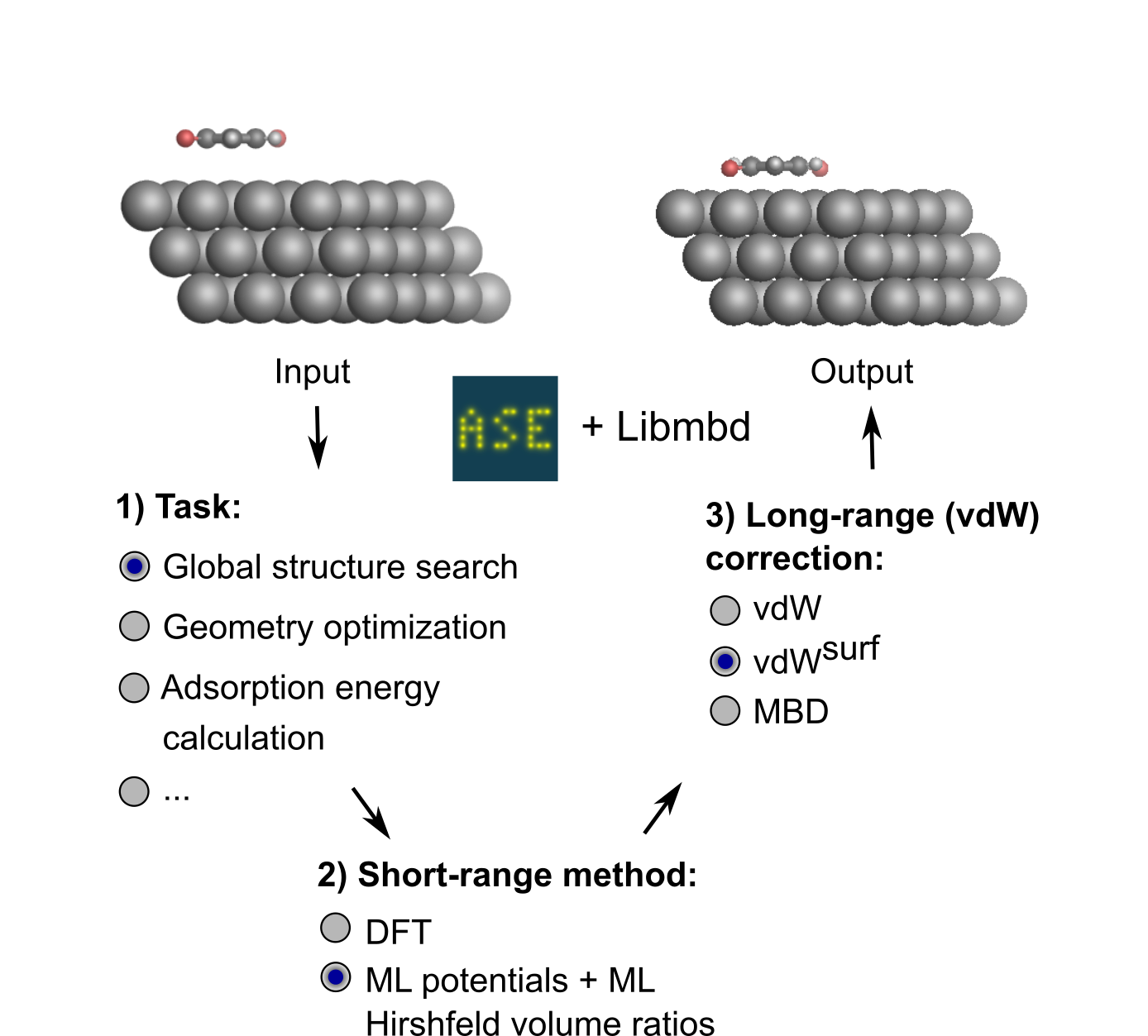}
    \caption{Overview of the method developed in this work. Different machine learning interatomic potentials (MLIPs) that allow for the computation of Hirshfeld volume ratios can be combined with different flavors of van der Waals (vdW) corrections, e.g. screened vdW pairwise interactions~\cite{Tkatchenko2009PRL} and many-body dispersion (MBD).~\cite{Tkatchenko2012PRL} The so-obtained MLIPs are interfaced with the Atomic Simulation Environment (ASE)~\cite{Larsen2017IOPP} and can be used for global structure searches, optimizations, energy predictions or other types of simulations implemented within ASE.}
    \label{fig:1}
\end{figure}

%This work
In this work, we present a deep learning approach to efficiently predict structures and stabilities at metal-organic interfaces for the purpose of high throughput structural (pre)screening and global energy landscape exploration. To this end, we create an approach that combines an NN-based MLIP with an established long-range dispersion method from the TS family of methods. As shown in Fig.~\ref{fig:1}, the short range description is provided by a local MLIP, whereas the long-range interaction is provided by one of the TS methods such as MBD. We couple the two approaches by constructing an ML representation of a partitioning of the electron density based on Hirshfeld atoms-in-molecules volumes.~\cite{HirshfeldCA1977,Tkatchenko2009PRL} This rescales atomic polarizabilities that enter the long-range description based on the local chemical environment of the atoms provided by the DFT description of short-range interactions. We deliver an open-access implementation of this approach by coupling the Atomic Simulation Environment (ASE) code~\cite{Larsen2017IOPP} with the Libmbd package.~\cite{libmbd} To further increase the robustness of our approach, we implement query-by-committee,~\cite{Freund1997ML,Melville2004,Behler2015IJQC} which establishes the model variance in energy and force predictions. This allows us to define a dynamic stopping criterion for when the prediction of the MLIP becomes unreliable and structure optimizations have to be continued with electronic structure theory. This is particularly useful in the context of efficient pre-relaxation of structures to reduce the computational cost associated with structure search. We show the utility of this approach on two systems, namely a global structure search for gold (Au) NCs adsorbed onto a diamond (110) surface and the structural relaxation of large conjugated organic molecules, namely 9,10-anthraquinone (A2O), 1,4-benzoquinone (B2O), and 6,13-pentacenequinone (P2O), summarized as X2O, adsorbed onto a silver (Ag) (111) surface that self-assemble into a variety of surface phases.~\cite{Jeindl2021ACSN} This method can be used to obtain optimized structures close to DFT minima with adsorption heights in good agreement to DFT. The model for X2O on Ag(111) is trained on sparse data extracted from open data repositories, which shows the utility of the model to facilitate structure pre-relaxations. We further demonstrate that the ML models trained on these data are transferable to different aromatic organic molecules on the same surface that were not contained in the training data set.

\section{Methods}

\subsection{ML potentials coupled to long-range dispersion corrections}
The TS vdW and MBD methods are \emph{a posteriori} corrections to DFT, although they both also exist as self-consistent variants.\cite{Ferri2015} Throughout this section, we refer to vdW, but note that the same arguments hold true for vdW$^{\textnormal{surf}}$.\cite{Ruiz2012PRL}
In the case of the vdW scheme, the dispersion energy contribution is a pairwise potential:\cite{Tkatchenko2009PRL}
\begin{equation}
    E_{\mathrm{vdW}}(\mathbf{R}) = - \sum_{\mathrm{A,B}}f(r_{\mathrm{cut}},A,B)\frac{C_6^{\mathrm{AB}}(\mathbf{R})}{{R_\mathrm{AB}}^6}
\end{equation}  
where $R_{\textnormal{AB}}$ is the distance between two atoms, A and B, and $f$ is a damping function to avoid double counting of short-range contributions. The model depends on tabulated free atom reference parameters such as atomic polarizabilities that are used to calculate $C_6^{\textnormal{AB}}$ coefficients and scaled vdW radii that define $r_{\textnormal{cut}}$ in the damping function. The $C_6^{\textnormal{AB}}$ coefficients explicitly depend on all coordinates of the system $\mathbf{R}$ to account for the chemical environment of the atoms. This is achieved by re-scaling the atomic polarizabilities and vdW radii based on the Hirshfeld atoms-in-molecules partitioning scheme.~\cite{HirshfeldCA1977} The ratio between effective volume of an atom in a molecule and a free atom is used as re-scaling factor:\cite{Tkatchenko2009PRL, Stoehr2016JCP}
\begin{equation}\label{eq:1}
    H_\mathrm{A} = \frac{V_{\mathrm{A,eff}}}{V_{\mathrm{A,free}}}.
\end{equation}

The MBD scheme is an extension of the vdW method that accounts for long-range electrostatic screening. This description is achieved by adding long-range screening effects to the effective atomic polarizabilities.

In this work, we couple both the vdW and MBD long-range dispersion schemes to an MLIP by creating an ML model of the Hirshfeld-based scaling ratios ($H_\textnormal{A}$) for all atoms A in the system. We note that the range-separation parameter in MBD and damping coefficient used in vdW are the only parameters specific to the employed exchange-correlation functional approximation to which the dispersion correction is coupled. As we train MLIPs to reproduce training data created with a specific exchange-correlation functional, we can retain the same parameters as used for the respective functional for vdW corrections to the generated MLIP.

%computational implementation
Throughout this work, we employ the ASE code which offers calculator interfaces to various electronic structure packages.~\cite{Larsen2017IOPP} The ML models in this work are based on the continuous-filter convolutional NN SchNet~\cite{Schutt2018,Schutt2017_double,Schuett2019JCTC}, which is a message-passing NN that learns the representation of the atomic environments in addition to its relation to the targeted output. ASE also provides an interface to the deep learning toolbox SchNetPack to employ NN-based MLIPs within ASE.\cite{Schuett2019JCTC} We have implemented an ASE calculator interface for the Libmbd code~\cite{libmbd} and further implemented an ASE calculator instance that combines a short-range calculator (e.g. electronic structure package or MLIP based on SchNetPack) with a Libmbd calculator instance. This interface calculator passes Hirshfeld scaling ratios predicted by an ML model into the Libmbd calculator to perform vdW- or MBD-corrected SchNet (denoted `ML+vdW' and `ML+MBD', respectively) calculations. All developed code is freely available on GitHub.~\cite{schnetvdW}

\subsection{Training Data}

\subsubsection{Gold Nanoclusters on Diamond (Au@C)}
DFT calculations were conducted using the all-electron numeric atomic orbital FHI-aims~\cite{aims} code and the Perdew-Burke-Ernzerhof (PBE)~\cite{Perdew1996PRL} exchange-correlation functional. The numeric atomic orbitals were represented using a `light' basis set and dispersion effects were accounted for via the MBD scheme.~\cite{Tkatchenko2012PRL} The total energy, sum of eigenvalues, charge density, and energy derivatives convergence criteria were set to $1\times10^{-6}$ eV, $1\times10^{-2}$ eV,  $1\times10^{-5}$ e/${a_0}^3$, and $1\times10^{-4}$ eV/\AA\, respectively. For structure relaxations, the maximum residual force component per atom was set to $1\times10^{-2}$ eV/\AA. Initial structures were constructed using ASE~\cite{Larsen2017IOPP} with Au NCs of various sizes adsorbed onto the center of a diamond (110) surface, with all carbon (C) atoms being fully frozen during optimizations. To lower computational costs and memory requirements, we create an aperiodic cluster cut-out of a diamond surface that corresponds to a $7\times7$ supercell repeat of a 7-layered diamond (110) slab. 
%Thus, strictly speaking, this model rather serves an example of a carbon support similar to boron-doped diamond than  diamond surface. Still, due to the large size of the surface and the facts that we froze all C atoms and placed Au NCs on the centre of the surface, the model can still be regarded a good approximation of a diamond surface. 
%We refer to this model to Au NCs on diamond in the following.  
An example of an Au NC  with n=50 (n denotes the number of Au atoms) on a diamond (110) surface can be seen in Fig. \ref{fig:scatter}d.

The starting point for the training dataset for Au@C models were 62 geometry optimizations of Au NCs on diamond (5, 4, 8, 8, 9, 10, and 18 geometry relaxations were conducted on Au clusters of size $n=$ 15, 20, 30, 35, 40, 45 and 50 atoms, respectively, on the aforementioned diamond (110) surface model). The training data points were collated using every relaxation step of the optimization runs, which therefore included both optimized and not fully-optimized structures. These computations led to an initial training dataset comprising 5,368 data points, which we used to train four MLIPs (trained on energy and forces). All MLIPs were trained using the same dataset, which was split randomly into training, validation, and test sets. All ML models trained on the initial training dataset are denoted as "ML$_\textnormal{init.}$". MLIPs were used to predict `local' energies and forces as well as Hirshfeld volume ratios to correct for long-range interactions at the MBD level. For energies and forces, we trained a set of models to use the query-by-committee approach discussed in subsection \ref{qbc}, which makes energy predictions more robust by a factor of $\sqrt{q}$, where $q$ is the number of trained ML models. The training process of energies and forces is explained in detail in section S1.1 in the SI. The models slightly differed in the weights of energies and forces used in the combined loss function (see equation 1 and discussion in the next subsection).
%One ML model to predict the Hirshfeld volume ratios was used for simulations. 
The model architecture and hyperparameter optimizations for the Hirshfeld model can be found in the SI in section S1.2.

To extend the training dataset, adaptive sampling~\cite{Behler2015IJQC} was carried out, which was originally developed for molecular dynamics simulations. Importantly, the predictions of the set of ML models are compared at every time step. Whenever the variance of the models exceeded a predefined threshold (with the threshold often being set slightly higher than the root-mean-squared error of the models on a test set\cite{Westermayr2019CS}), the data point was deemed untrustworthy and recomputed with the reference method. This data point was then be added to the training set and the models retrained. In this work, we applied this concept to a global structure search using the basin-hopping algorithm~\cite{Wales1997JPCA,Wales1999S} as implemented in ASE~\cite{Larsen2017IOPP} rather than MD simulations. After each geometry optimization during the basin-hopping run, the variance of the model predictions was computed and geometries with the largest model variances were selected for further DFT optimizations. These optimizations were then added to the training set. Stopping criteria for ML optimizations are discussed in section \ref{qbc}.

In total, three adaptive sampling runs were carried out. The first adaptive sampling run was carried out with the initial ML models, "ML$_\textnormal{init.}$". After data points were sampled and the dataset was extended, ML models were retrained. MLIPs after the first adaptive sampling run (denoted as ML$_\textnormal{adapt.1}$) were trained on 7,700 data points for training and 800 data points for validation. With these models, the second adaptive sampling run ML$_\textnormal{adapt.2}$ was executed. A total of 9,757 data points were collected after the second adaptive sampling run. ML$_\textnormal{adapt.2}$ models were trained on 8,500 data points for training and 800 data points for validation. After the final adaptive sampling run (ML$_\textnormal{adapt.3}$), there were a total of 15,293 data points. 12,500 data points were used for training and 1,500 for validation. More details on the adaptive sampling runs can be found in section S1.1. 

\subsubsection{Organic Molecules on Silver (X2O@Ag)}\label{x2oml}
The training data points for X2O@Ag are taken from the NOMAD repository~\cite{A2O,B2O,P2O} and are based on Ref. \citenum{Jeindl2021ACSN}. X2O summarizes different functional organic monomers, which are described as monolayers on Ag(111) surfaces (abbreviated as X2O@Ag). As mentioned above, the three different molecules tested were: 9,10-anthraquinone (A2O), 1,4-benzoquinone (B2O), and 6,13-pentacenequinone (P2O) as shown in Fig.~\ref{fig:scatter}h. The dataset consists of 8,202 data points, where each data point comprises a geometry and the corresponding energies, forces, and Hirshfeld volume ratios. In more detail, the datasets contain 353 data points of the clean substrate in total (about 4\% of the data), 1,397 data points of P2O molecules, 2,249 data points of A2O molecules, and 4,156 data points of B2O molecules. The molecules were either in the gas phase, arranged as two-dimensional free-standing overlayers in various unit cells and arrangements (5,724 data points; about 70\% of the data), or adsorbed onto an 8-layered Ag(111) surface slab (2,125 data points; about 26\% of the data). Some supercells contained several different molecules adsorbed onto the surface. The reference data points possessed different unit cell sizes and the reference method for the data was vdW$^{\textnormal{surf}}$-corrected DFT (DFT+vdW$^{\textnormal{surf}}$) with the PBE exchange-correlation functional, with a dipole correction also being employed. A `tight' basis set was used for the top three substrate layers while a `very light' basis set was used for the five lower lying layers.~\cite{Jeindl2021ACSN}. The data points were taken from 208 geometry relaxations and 6,773 single-point calculations.
The training set data was generated with FHI-aims in ref.~\citenum{Jeindl2021ACSN}, with the total energy, forces, and charge density convergence criteria were set to $1\times10^{-5}$ eV, $1\times10^{-3}$ eV, $1\times10^{-2}$ $e/a_0^3$, respectively.

For Au@C, four ML models were trained on energies and forces (see section S1.1 for details) and one model on Hirshfeld volume ratios, which was used in all geometry optimizations. As mentioned earlier, adaptive sampling was not carried out for this dataset as we wanted to base our models purely on sparse existing data derived from a small set of geometry optimizations to showcase the usability of our model to speed up structure relaxations.

In addition, both DFT and ML structure relaxations of 16 B2O@Ag systems far away from the surface were conducted and served as a test set. These structures are especially challenging to relax as common optimization algorithms often fail for systems that are far away from the optimized structure, even with DFT and long-range interactions. One problem is that vdW forces decrease quickly with the distance of an adsorbate to the surface, and quasi-Newton optimizers with simple Hessian guesses can converge to a geometry that has hardly changed compared to the initial structure. % Challenging systems are for instance adsorbates far away from the surface or systems that stand upright when they should be flat according to the optimized structure.
This problem can be overcome by using an improved Hessian approximation for the initialization of the optimization. In this work, we used the Lindh Hessian~\cite{Lindh1995CPL,aims} to initialize structure relaxations for DFT+vdW$^{\textnormal{surf}}$ and ML+vdW$^{\textnormal{surf}}$ calculations. The same optimization criteria were used as in the reference calculations, but we used the ASE calculator with our vdW implementation rather than FHI-aims for consistency. 

\subsection{Machine Learning Interaction Potentials (MLIPs)}\label{mlsec}
We generate vdW-free SchNet\cite{Schuett2019JCTC,Schutt2017_double} MLIPs and a SchNet-based model for the Hirshfeld volume ratios. The local vdW-free potential energy surfaces were obtained by subtracting the vdW corrections from the total energies and forces obtained with FHI-aims.
%The PESs modeled by SchNet are obtained as the sum of atomic contributions, where every atom is modeled in its chemical and structural environment within a cutoff region. 
The MLIPs are trained with vdW-free energies ($E$) and forces ($F$). The forces are treated as derivatives of the MLIP, $E_{\textnormal{local}}^{\textnormal{ML}}$, with respect to the atomic positions ($\mathbf{R}$) and are trained in addition to the energies using a combined loss function ($L_2$):
\begin{equation}\label{eq:l2}
\begin{array}{ll}
      L_2 = t\left|\left| E_{\mathrm{local}}^{\mathrm{QC}} - E_{\mathrm{local}}^{\mathrm{ML}} \right|\right|^2 + (1-t) \left|\left| F_{\mathrm{local}}^{\mathrm{QC}}-\frac{\partial E_{\mathrm{local}}^{\mathrm{ML}}}{\partial \mathbf{R}}\right|\right|^2,  \\
      \mathrm{where~} E_{\mathrm{local}}^{\mathrm{ML}} = \displaystyle \sum_\mathrm{A}^{N} E_{\mathrm{local,A}}^{\mathrm{ML}}
\end{array}{}
\end{equation}{}
The energies are obtained as the sum of atomic contributions with $N$ being the total number of atoms in a system. The trade-off, $t$, is used to ensure a good balance between energies and forces during training.

In contrast, the Hirshfeld volume ratios were fitted per atom using another SchNet model that was adapted for this purpose. The corresponding loss function, $L_2^H$:
\begin{equation}\label{eq_l2h}
    L_2^{H} = \sum_\mathrm{A}^{N} \left|\left| H_\mathrm{A}^{\mathrm{QC}}-H_\mathrm{A}^{\mathrm{ML}} \right|\right|^2,
\end{equation}
contains all Hirshfeld volume ratios, allowing for all values to be modeled in one atomistic ML model. The details on the model training and the used parameters for model training can be found in the SI in section S1.2. %Model errors for each ML model are summarized in Table S1 and final models are shown using scatter plots in Fig.~\ref{fig:scatter}(a). Geometry relaxations with MLIPs and the criteria when to stop an optimization are discussed in the next section.

%\section{Hybrid nanosystem}
As mentioned in the previous subsection \ref{x2oml} the X2O@Ag data was generated using two basis sets for Ag atoms depending on their position. Different basis sets will result in different energies and forces. Therefore, the dataset was pre-processed prior to training by representing all the Ag atoms that were described using a `very light' basis set with a different atom label. This process allowed the MLIPs to be trained on data with mixed basis sets.

\subsection{Structure Relaxations with MLIPs}\label{qbc}
\begin{figure*}[ht]
    \centering
    \includegraphics[scale=0.85]{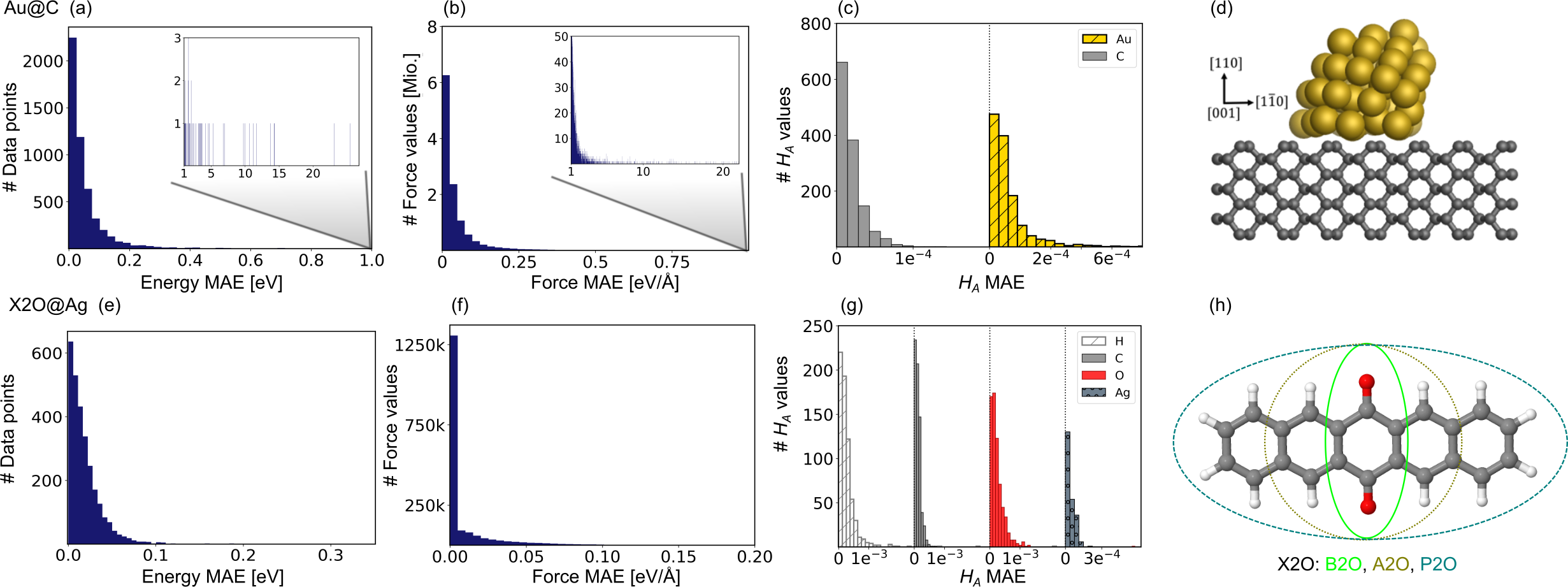}
    \caption{Prediction errors for gold nanoclusters (NCs) on diamond (110) surfaces (Au@C) on top and for X2O systems on Ag(111) (X2O@Ag) in the bottom. (a,e) Mean absolute errors (MAEs) for energies, (b,f) for forces (middle), and (c,g) Hirshfeld volume ratios, $H_\textnormal{A}$, for Au@C and X2O@Ag, respectively. Bar plots for energies and forces are shown and summarized from four trained machine learning (ML) models. For forces, the error with respect to each force component is shown, i.e., one data point thus contains as many components as thrice the number of atoms (around 2,100 values for Au@C and about 200-300 for X2O@Ag systems) for the three orthogonal directions, which are [110], [001] and [1$\overline{1}$0] for Au@C, and [111], [1$\overline{2}$1] and [$\overline{1}$01] for X2O@Ag. For Hirshfeld volume ratios, one ML model is used, and the error is split into contributions from the separate atom types. (d) Example of an Au NC with 50 atoms on a diamond (110) surface and (h) X2O systems in the gas phase that are described in this study on Ag(111).}
    \label{fig:scatter}
\end{figure*}

For all structure relaxations, local MLIPs and ML Hirshfeld volume ratios were used for additional vdW corrections, and the screened atomic polarizabilities suggested for Ag by \citet{Ruiz2012PRL} were used to account for the correct dielectric screening of the metal surface.
Structure relaxations were carried out using the Broyden–Fletcher–Goldfarb–Shanno (BFGS) algorithm, as implemented in ASE,~\cite{Larsen2017IOPP} which utilized a maximum atomic force criterion, $fmax$, to decide when the optimization should be stopped. We adopted the decision as to when the optimization should be stopped by further making use of the query-by-committee concept and taking the variance of the ML model predictions for energies into account.%; a detailed discussion can be found in the supplementary information (SI) in section S3.1, hence we only briefly describe the process here.

The query-by-committee approach~\cite{Freund1997ML,Melville2004,Behler2015IJQC} takes the mean of the predictions of $q$ ML models for a given property, $P$: $P^{\mathrm{ML}}=\frac{1}{q}\displaystyle\sum_{i=1}^q P^{\mathrm{ML}_q}$. In all subsequent calculations, we follow the mean of the potential energy surface and corresponding forces. 
While the accuracy and robustness of the predictions can be improved by a factor of $\sqrt{q}$,~\cite{Gastegger2017CS}
%when data is sparse, as shown recently in Ref.~\citenum{Gastegger2017CS} for ML potential energy surfaces. 
no improvement for the predictive accuracy of other properties such as dipole moments, could be achieved.
We also found that the prediction of Hirshfeld volume ratios was not improved by the query-by-committee approach, so only one ML model was used for learning Hirshfeld volume ratios in the following. The reason for this can be manifold and is likely due to the fact that the accuracy of the Hirshfeld volume ratio models is already very high as compared to the energy models, which is why query-by-committee is unlikely to strongly improve the prediction accuracy of Hirshfeld volume ratios.

A further consequence of having more than one ML model for energies is that this approach allows us to assess the reliability of the ML predictions by computing the model variances,
\begin{equation}\label{eq:var}
E_{\mathrm{var}}^{\mathrm{ML}} = \sqrt{\frac{1}{1-q}\sum_{i=1}^{q}(P^{\mathrm{ML}_q}-P^{\mathrm{ML}})^2}.
\end{equation}
The assessment of the reliability of predictions is especially important when ML models serve as pre-optimizers and cannot reliably reach a low $fmax$ value.

To find optimal stopping criteria of the optimization with ML models, we explored a random grid of 1,000 different stopping criterion combinations for structure relaxations of the Au@C test set using ML$_\textnormal{init.}$ and the X2O@Ag test set (see Fig. S2 a and b, respectively). The ability to perform 1,000 geometry optimizations as a test further showcases the computational efficiency of the approach. Test runs showed that introducing an additional initial $fmax_\textnormal{init.}$ value as a threshold, after which the ML model variance for energies, $E_{\textnormal{var}}^{\textnormal{ML}}$ (eq. \ref{eq:var}) is monitored, is beneficial with respect to the agreement of the final ML-optimized structure and DFT-optimized structure. The $fmax_\textnormal{init.}$ value was found to be relatively robust and set to 0.15\,eV/{\AA} for the test studies shown in this work, but it can be set to a different value by the user to take into account the reliability of ML models. %Almost identical results were obtained when using a different value between 0.1-0.2\,eV/{\AA}.

As soon as the $fmax_\textnormal{init.}$ value was reached during an optimization, the number of consecutive steps that showed rising energy variances was monitored.
The amount of consecutive steps that showed rising energy variance was varied in a grid search and we found three consecutive steps of increasing energy variances to be a good criterion to stop the optimization algorithm with final structures closest to the DFT reference minimum (Fig. S1). The energy variance between different ML models will always fluctuate around a small number, even in the case of reliable geometry relaxations. Hence, the energy variance can become larger in consecutive steps without necessarily indicating that the structure relaxation becomes unreliable. Three consecutive steps in which the energy variance was rising was found to be small enough to still ensure that the structure is not already too far away from the last reliable structure. 
To further ensure that the optimization did not run out of the training regime, we terminate the algorithm after $fmax_\textnormal{init.}$ was reached and after that,  whenever the model energy variance reached a high value that we set to 1\,eV or when the $fmax$ jumped to a value that was larger than 2\,eV/{\AA}. Both events were observed when model predictions ran into regions not supported by training data. 
For ML$_\textnormal{adapt.3}$ models, an $fmax$ value of 0.05\,eV/{\AA} was able to be reached, hence the additional stopping criteria were not required using these refined models.

\section{Results}
\subsection{Model Performance}\label{sec:modelperformance}

Fig. \ref{fig:scatter} shows model prediction errors for the vdW-free MLIPs for energies and forces and the Hirshfeld ratio ML models in panels a, b, and c respectively for Au@C and panels e, f, g, respectively, for X2O@Ag models. The mean absolute errors (MAEs) and root-mean-square errors (RMSEs) on the data points of the hold-out test set shown in Fig. \ref{fig:scatter} for energies, forces, and Hirshfeld volume ratios can be found in Table S1 in the SI.

The MAE of the four models ranges from 0.017 to 0.021 eV for energies and 0.021-0.025 eV/{\AA} for forces for X2O@Ag. ML models trained on Au@C have MAEs of 0.013 to 0.18 eV for energies and 0.014 to 0.26 eV/{\AA} for forces. As can be seen, there are some outliers in the data set of Au@C with errors on these data points shown in the insets of top panels a and b. These data points are geometries with unfavorable structures and energies far out of the region in which most data points lie. These data points were included to ensure that the model was able to rank structures correctly and predict energetically unfavorable structures with high energies. %avoid certain regions during optimizations. 
For training on these data points, the $L_2$ loss was adapted to a smooth version of the $L_1$ loss, which is explained and defined in section S1.2.
%ordering b2o[13 12  5 10 11  6  8  0  7  3 14  1  2  4 15  9]

Besides data points representing unfavorable Au@C NCs with large vdW-free energies and vdW-free forces that were explicitly introduced into the training set, the ML models predict vdW-free energies, vdW-free forces, and Hirshfeld volume ratios accurately. The MAE for the Hirshfeld volume ratios, a quantity that ranges between about 0.6 and 1.05, is $3.9\times 10^{-4}$ and  $1.1\times10^{-4}$ for X2O@Ag and Au@C, respectively. 

In the following, we will assess the performance of the proposed method by performing structure relaxations of geometries of two additional hold-out test sets for X2O@Ag and Au@C. These hold-out test sets comprise full structure optimizations and none of the geometry optimization steps during the relaxations were included for training.

\subsection{Global Structure Search: Gold Nanoclusters on Diamond (Au@C)}
\begin{figure}[ht]
    \centering
    \includegraphics[width=3.3in]{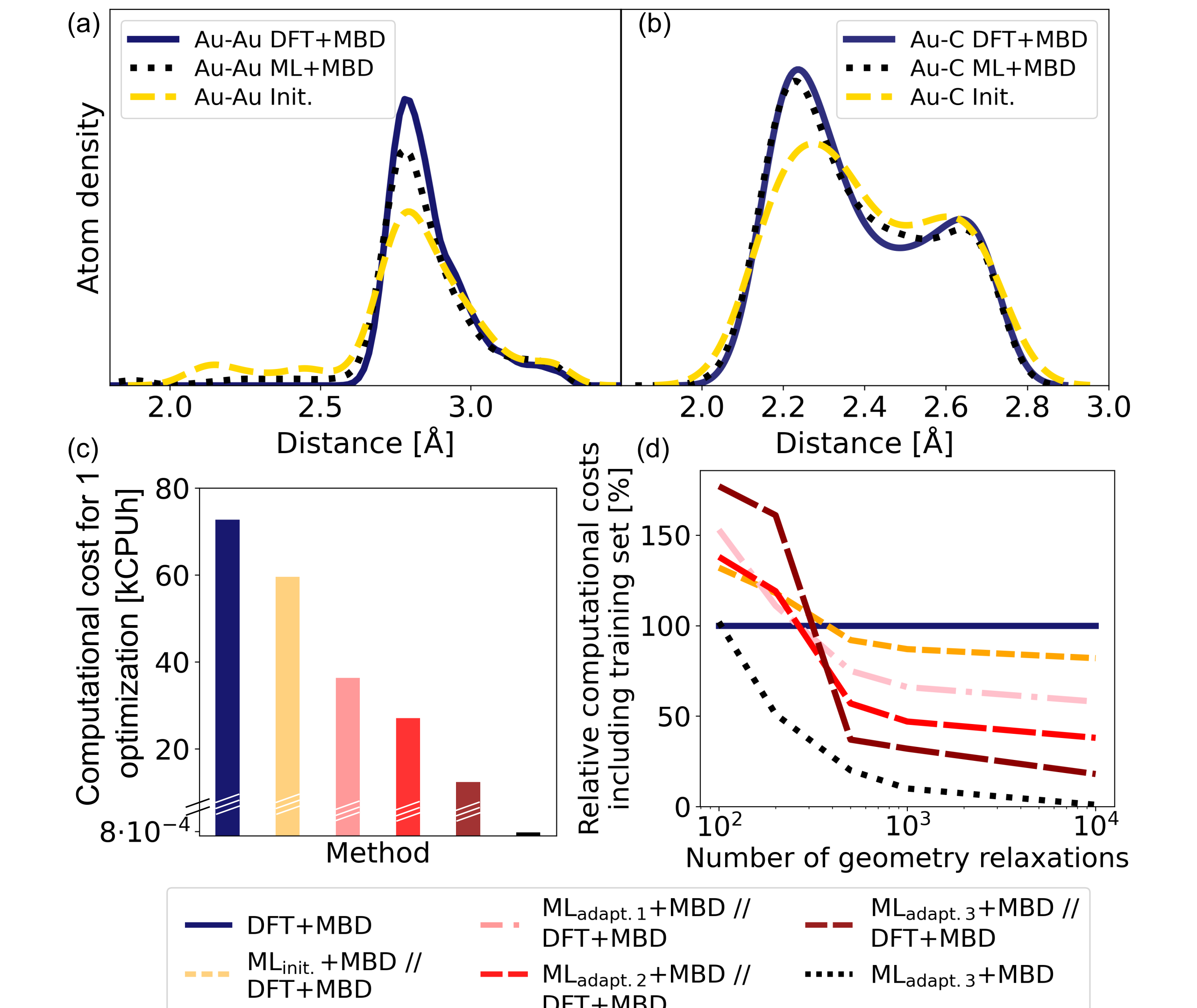}
        \caption{(a) Kernel density estimate for the radial atom distribution of Au-Au and (b) Au-C bonds of Au@C systems for the optimized structures with DFT that comprise the training set and were computed with DFT+MBD (solid lines, denoted DFT+MBD). The starting structures for geometry optimizations are denoted using "Init."  and dashed lines and the ML+MBD-optimized (ML$_\textnormal{adapt.3}$+MBD) structures are shown in dotted lines. (c) Computational costs in kilo central processing unit hours (kCPUh) of a single Au@C structure relaxation performed with DFT+MBD (blue), and prerelaxations with ML+MBD models followed by further optimization with DFT+MBD (denoted ML+MBD//DFT+MBD). (d) Computational cost including model training cost as a function of the number of performed geometry relaxations. Computational costs were assessed by defining an average time per geometry optimization that was based on the initial training data. }
    \label{fig:auc1}
\end{figure}

As NCs can exhibit many metastable geometries, we first assess the performance of our model with respect to interatomic distances and then evaluate the applicability of our approach to energetically differentiate between different cluster geometries. For the first task, we use a test set of Au@C models that contain DFT+MBD optimizations of Au NCs on diamond (110) with cluster sizes of $n$ = 6, 15, 20, 25, 28, 30, 35, 40, 44, 45, 60, and 66. On average, 95 optimization steps were required with DFT+MBD for one geometry optimization. All initial starting structures for geometry optimizations of NCs were created with ASE, where the NCs were placed onto the center of a diamond (110) surface. The same starting geometries as used in DFT structure optimizations were taken for structure relaxations with the final model obtained after the third adaptive sampling run, denoted ML$_\textnormal{adapt.3}$+MBD. The minima found with ML$_\textnormal{adapt.3}$+MBD were assessed according to the radial atom distributions of the Au NCs in Figure \ref{fig:auc1}a. Radial atom distributions obtained from structures using the ML$_\textnormal{adapt.3}$+MBD scheme are similar to those from DFT+MBD. For the Au-Au radial atomic distribution in panel a, distances at values smaller than around 2.6 {\AA} are removed by geometry optimization and the main distance distribution at around 2.8 {\AA} aligns well with DFT+MBD. Slight deviations can be found at 2.5 {\AA} for Au-C in panel b, which can also be seen in the radial atom distributions for the starting structures used for geometry optimizations (denoted  as "init."). The peaks of the initial distribution are shifted towards the DFT+MBD peaks upon optimization. 

\begin{figure*}[ht]
    \centering
    \includegraphics[width=7in]{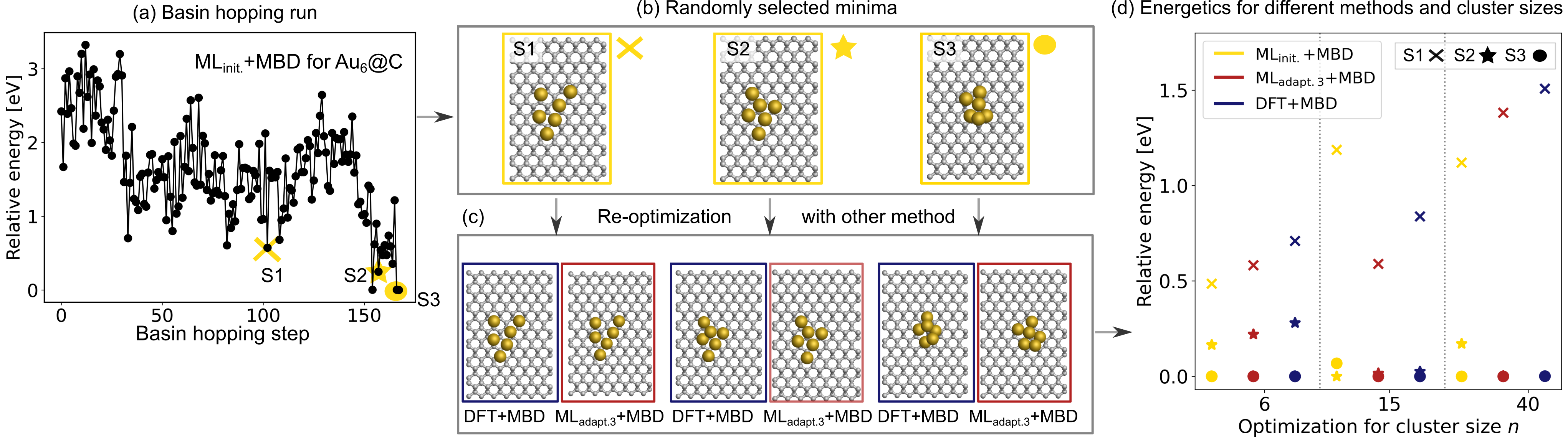}
    \caption{ (a) Basin hopping run with ML$_\textnormal{init.}$ for Au@C with Au$_6$ nanoclusters (NCs). Yellow circles indicate (b) 3 selected structures S1-S3 that include the energetically lowest geometry and two randomly selected structures according to ML$_\textnormal{init.}$ that are (c) reoptimized with DFT+MBD (blue) and ML$_\textnormal{adapt.3}$+MBD (red). (d) Relative energies reported with respect to the energetically lowest cluster for each method. In addition, energy ranking of the energetically lowest structures and two randomly selected structures from basin hopping runs with NC sizes of 15 and 40 atoms using ML$_\textnormal{init.}$+MBD (yellow), ML$_\textnormal{adapt.3}$+MBD (red), and DFT+MBD (blue). Corresponding structures are shown for each method in Fig. S2.}\label{fig:AuC}
\end{figure*}

The benefit of using ML+MBD instead of DFT+MBD lies in the reduction of computational effort associated with structure relaxations. 
Figures \ref{fig:auc1}c and d show the computational costs of structure relaxations with ML+MBD, DFT+MBD and a ML+MBD pre-optimization followed by a DFT+MBD optimization (denoted `ML+MBD//DFT+MBD'). Panel c shows the cost of a single structure relaxation in kilo-central processing unit hours (kCPUh), recorded on dual AMD EPYC$^\textnormal{TM}$ Zen2 7742 64-core processors at 2.25 GHz. As can be seen, the computational cost of ML+MBD optimization (black) is about 0.01\% of the cost of DFT+MBD. 
However, it can be argued that the structure relaxations solely conducted with ML+MBD might not be accurate enough for a specific purpose and are not sufficiently close to DFT+MBD. To this aim, we performed DFT+MBD optimizations using the optimized structures obtained from the ML$_\textnormal{init.}$ (yellow), ML$_\textnormal{adapt.1}$ (pink), and ML$_\textnormal{adapt.2}$ (red), and ML$_\textnormal{adapt.3}$ (dark red) models and summed up the computational expenses from respective ML+MBD and additional DFT+MBD calculations. In this approach, ML+MBD acts as a pre-optimization method. As expected, the computational cost increases when combining ML+MBD with DFT+MBD. However, the better the optimized structure resulting from the ML model, the fewer DFT+MBD optimization steps are required. This is why the combination of refined adaptive models with DFT require less computational cost for the same task than the initial model in combination with DFT.

Fig. \ref{fig:auc1}d plots the computational cost of performing one to 10,000 structure optimizations of the different models including the cost of generating the training data set for the ML model construction. The costs are extrapolated and are shown relative to DFT+MBD (100\%, dark blue). As can be seen from the dotted black lines, using the final ML model, ML$_\textnormal{adapt.3}$+MBD can greatly reduce the computational costs whilst still achieving good accuracy (see panels a and b). Note that ML+MBD values include the cost of training data generation and model training. In case of large scale screening studies, where many geometry optimizations are required, it is clearly beneficial to use refined and accurate ML+MBD models. In cases where high accuracy is required, a subsequent re-optimization with DFT+MBD to reach an $fmax$ of 0.01 eV/{\AA} may be necessary. In this scenario, we find that the ML+MBD//DFT+MBD optimization sequence is only computationally beneficial to standalone DFT+MBD optimization if the number of required structural relaxations is between 100 and 500. In Fig.~\ref{fig:auc1}d, $\textnormal{ML}_\textnormal{init.}-\textnormal{ML}_\textnormal{adapt.3}$ refers to models trained on more and more data points. The break-even point in terms of computational cost for ML+MBD//DFT+MBD is similar for all models, but lowest for "adapt.2" (about 100 structure relaxations) and highest for "init." (about 500 structure relaxations). This shows that there is a sweet spot for the construction of MLIPs between the cost of creating an (overly) large training data set and the computational time saving benefit.

To validate the reliability of the structure and stability prediction of the ML+MBD models for Au@C, three basin-hopping optimization runs that were carried out for the initial adaptive sampling runs for clusters of size $n=$ 6, 15 and 40  were selected. The global minimum and two random local minima were selected from each basin-hopping run for the different cluster sizes. 
The basin-hopping run for a cluster size of $n$ = 6 is shown in Fig.~\ref{fig:AuC}a. The three structures used for validation are denoted S1$-$S3 (yellow in panel b) and were re-optimized with DFT+MBD (blue) and ML$_\textnormal{adapt.3}$ (red) separately. In panel Fig.~\ref{fig:AuC}c, the structures of DFT+MBD are compared to those of ML$_\textnormal{adapt.3}$+MBD. The structures are very similar to each other with slight deviations visible in geometry S3.

\begin{figure*}[ht]
    \centering
    \includegraphics[width=5in]{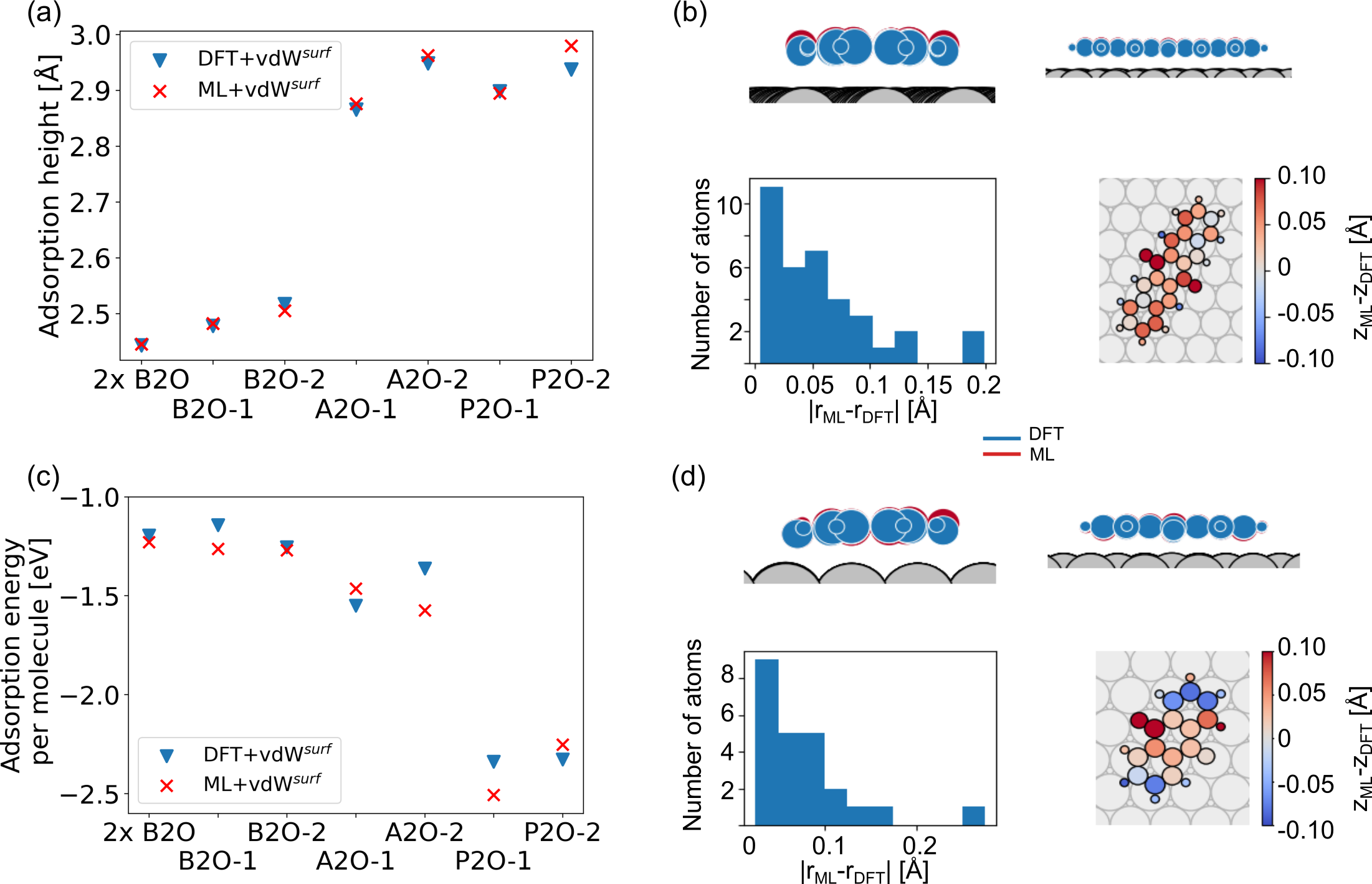}
    \caption{(a) Adsorption heights (average heights of all atoms in the molecule compared to the average heights of the first Ag layer) and (c) adsorption energies of X2O@Ag of a hold-out test set computed with DFT+vdW$^{\text{surf}}$ and ML+vdW$^{\text{surf}}$. The structures are single B2O molecules and two B2O molecules in a unit cell (denoted as "2$\times$B2O"), A2O, and P$_2$O on Ag(111) that differ in adsorption sites and orientation. %(Fig.s S3-S4) %and are shown from the front and bottom view in Fig.s S3 and S4, respectively. 
    (b,d) ML+vdW$^{\textnormal{surf}}$ structures (P2O-2 and A2O-2) compared to DFT+vdW$^{\textnormal{surf}}$ structures of panels (a) and (c).}
    \label{fig:x2o_main}
\end{figure*}

The energies of the three structures are plotted in Fig.~\ref{fig:AuC}d relative to the most stable structure. Even though the structures are not exactly the same, the energies are ranked similarly to each other. The ordering of the three structures is also correctly predicted with each method. As expected, the energy ranking of ML$_\textnormal{adapt.3}$+MBD is closer to the relative energy ordering of DFT+MBD than the initial ML model.
%Panels c and 
Panel d further shows the results of the same procedure carried out for cluster sizes of $n=$ 15 and 40, respectively. The structures for all clusters as predicted by all methods are visualized in Fig. S2 of the ESI. As can be seen, for the Au NC with 15 atoms, the energies are ordered incorrectly according to the initial model. The correct ordering of energies is established with the final model, ML$_\textnormal{adapt.3}$+MBD, and is similar to DFT. However, the highest energy geometry is predicted to be more stable than in the reference. This result could be an indication that the least favorable structure with a size of 15 is in a region of the potential energy surface that is under-represented in the training set. Indeed, the energy variance according to the query-by-committee approach is 4 times higher for this structure (around 30 meV) than for the other clusters (around 7 meV). For the Au NC with 40 atoms, the initial model suggested three energetically different structures, while the ML$_\textnormal{adapt.3}$+MBD and DFT+MBD methods suggest that the first two structures are identical in their energy. 
To conclude, ML combined with a long-range dispersion correction (MBD in this case) has proven powerful to reduce the costs of structure relaxations with DFT+MBD substantially. Given the rich diversity of structures and cluster sizes and the relatively few data points required, the model can be utilized as a pre-optimizer that leads to radial atom distributions close to the DFT+MBD optimum and can facilitate fast global structure searches including an approximate energy ranking of structures.

\subsection{Adsorption of Organic Molecules on Ag(111)}

\begin{figure}[ht]
    \centering
    \includegraphics[width=3.3in]{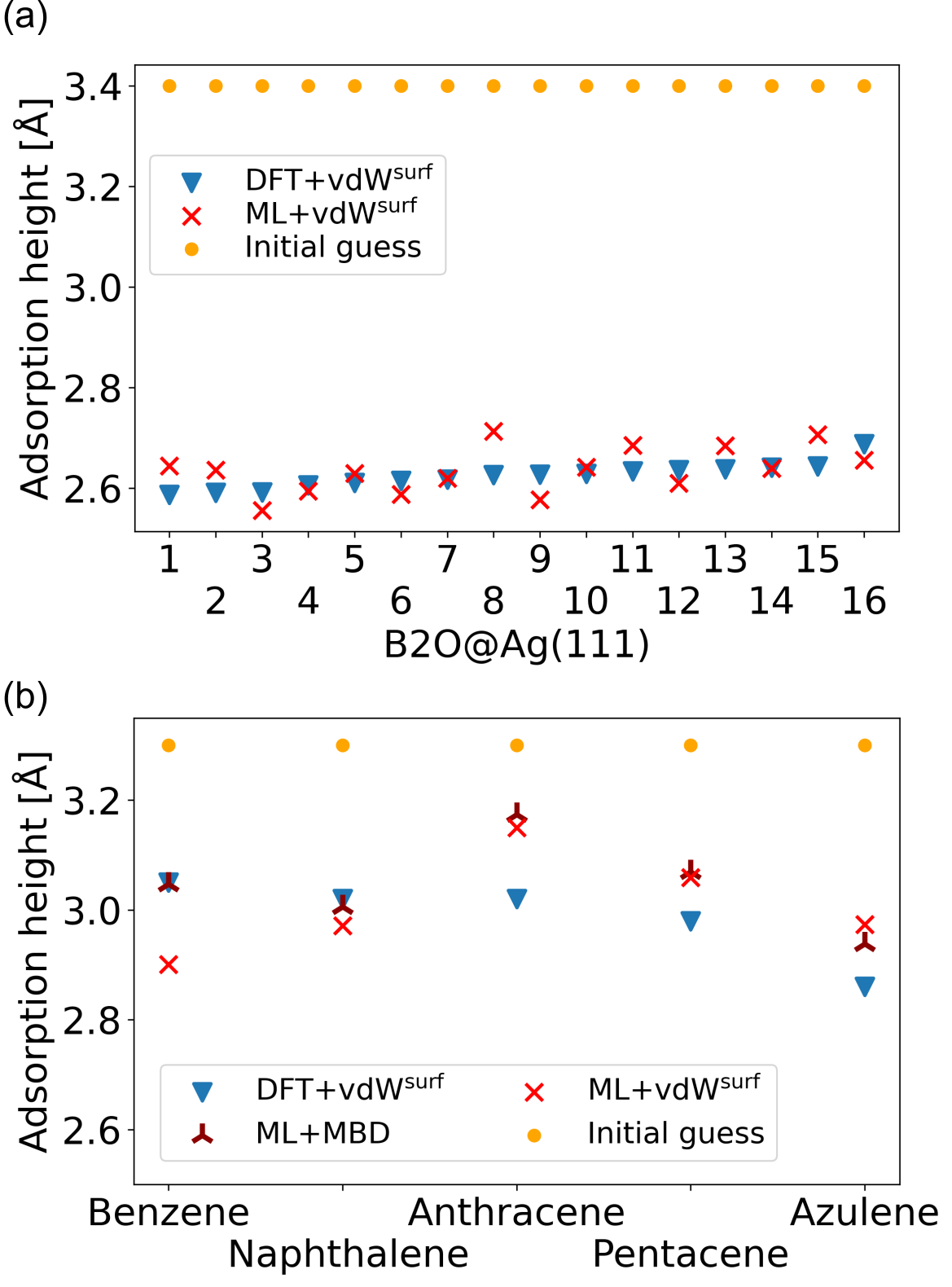}
    \caption{(a) Adsorption heights of B2O molecules on Ag(111).
    (b) Adsorption heights of benzene,~\cite{benzene_e_h}, naphthalene,~\cite{naphthalene_azulene_adsh} anthracene,~\cite{anthracene}% tetracene,~\cite{tetracene}, %removed it since there are no comparable values.
    pentacene,~\cite{pentacene_h}, and azulene,~\cite{naphthalene_azulene_adsh} computed with ML+vdW$^{\textnormal{surf}}$ and compared to DFT+vdW$^{\textnormal{surf}}$. The same adsorption sites as mentioned in the cited references (Table 1) are used.}
    \label{fig:x2o_transferability}
\end{figure}
Our second application case is based on organic molecules of the X2O family\cite{Jeindl2021ACSN} on Ag(111), as shown in Fig. \ref{fig:scatter}h. The existing training data set only includes few data points based on a small set of local geometry optimizations. We have defined a test set that contains randomly selected optimized structures held out from the training set. We removed several full structure optimizations, i.e., the starting geometries, the intermediate steps and the final optimized structures, from the training set to ensure no structure relevant for the test set is explicitly known by the models. The test set represents a small set of exemplary local minima of X2O molecules on a Ag(111) surface. The structures in the test set are denoted based on the type of organic molecule that is adsorbed on the surface, i.e., B2O, A2O, and P2O. The indices after the molecule abbreviations indicate geometries that differ in their adsorption site, orientation or cell size. One test example shows a unit cell with two B2O molecules. Fig.~\ref{fig:x2o_main}a and c show the adsorption heights and adsorption energies, respectively, of the ML+vdW$^{\textnormal{surf}}$-relaxed structures compared to the DFT+vdW$^{\textnormal{surf}}$-relaxed structures. The adsorption energies were obtained using the ML+vdW$^{\textnormal{surf}}$ method and reference adsorption energies were obtained from the DFT+vdW$^{\textnormal{surf}}$-optimized structures. Hence the energies in panel c are not obtained from identical geometries, but from the respective minimum energy structures of the methods. The adsorption energy is defined as $E_\textnormal{ads+Ag}-E_\textnormal{ads}-E_\textnormal{Ag}$, with ``ads'' referring to the adsorbate and ``Ag'' to the metal surface. Relaxed geometries of the clean surface and the isolated molecule were used as references in the calculation of the adsorption energy, and a negative adsorption energy value corresponds to an exothermic process. Adsorption heights were computed as distances of the average heights of the first Ag layer and the average heights of all atoms in the molecule.

The test to validate the new method is carried out as follows: the same starting geometries were used for ML+vdW$^{\textnormal{surf}}$ geometry relaxations as were used in DFT+vdW$^{\textnormal{surf}}$ reference optimizations. As can be seen from Fig. \ref{fig:AuC}a, our method reports adsorption heights that are very similar to those obtained with DFT+vdW$^{\textnormal{surf}}$. The structural similarity can be further assessed from panels b (P2O-2) and d (A2O-2), which shows the ML+vdW$^{\textnormal{surf}}$ compared to DFT+vdW$^{\textnormal{surf}}$ structures with the worst agreement in adsorption heights between ML and DFT. The top images show ML+vdW$^{\textnormal{surf}}$-optimized structures in red and DFT+vdW$^{\textnormal{surf}}$-optimized structures in blue. Bottom images show the error of each atom in \AA. The ML-predicted minimum energy structures are typically relatively close DFT predicted structures with the largest deviations in adsorption height per atom at about 0.2 \AA. Most deviations are below 0.05 \AA. Noticeably, these are not differences in bond lengths (Fig. S4) but absolute positions in z direction. Visualizations for the remaining structures presented in \ref{fig:x2o_main}a and c are shown in Fig. S3 of the ESI. 

In addition to the adsorption heights, we sought to assess the adsorption energies for the purpose of relative energy predictions of adsorption phases with respect to each other. As can be seen from panel c, the trend observed in the reference data can mostly be reproduced when comparing different molecules. There is hardly any trend in over- or underestimation of adsorption energies and the mean error on adsorption energies is around 0.10 $\pm$ 0.06 eV. 

As a more difficult challenge for the model, we generated an additional test set of 16 B2O structures on Ag(111) with DFT+vdW$^\textnormal{surf}$, which are far from the surface. These structures required around five to six times more optimization steps than the calculations in the training set and thus provide a test with initial structures that are much less favorable than those in the training set and the structures tested before. As mentioned briefly in the Methods section \ref{sec:modelperformance}, geometry optimization algorithms struggle with geometries far away from the surface and require additional considerations. To counter this problem, a two-fold optimization was conducted with our method. First, all atomic positions of the molecule were fixed apart from motion along the [111] direction, with the Ag(111) substrate fully constrained. After this initial relaxation, the molecule was allowed to relax into all directions and the top three Ag layers, as in the reference \citenum{Jeindl2021ACSN}, were also allowed to relax. To initialize the optimizations, we used the Lindh-Hessian\cite{Lindh1995CPL,aims} as was done in DFT+vdW$^\textnormal{surf}$ optimizations.
The results are shown in Fig \ref{fig:x2o_transferability}a. Our model gives fair adsorption heights for the systems when compared to the DFT reference and can be used as a computationally efficient pre-relaxation procedure without ever learning from data of  systems with large molecule-metal separation, as those were accounted for by the long-range dispersion correction. The mean error for adsorption heights is relatively low and around 0.04$\pm$0.02 {\AA}. 

The final challenge was to test our model for transferability to other organic molecules that have not been seen by the model. This would open the possibility to generate a fully transferable MLIP for hybrid metal-organic interfaces to be applied as a general structural pre-relaxation tool.
We test our approach on several different organic molecules adsorbed on Ag(111) that have been experimentally and computationally characterized previously, namely benzene, naphthalene, anthracene, pentacene (all from the acene family), and azulene. According to literature,~\cite{benzenesymm,naphthalene_azulene_adsh,anthracene,pentacene_ener} the most stable symmetry site was selected (indicated in table \ref{tab:X2O_extrapolation} in the first column). The gas-phase optimized structure of each organic molecule was placed around 3.3 {\AA} away from the surface. A similar two-step optimization procedure was applied as before. 
As shown in Figure \ref{fig:x2o_transferability}b, the trend in adsorption heights across molecules that is found with DFT+vdW$^\textnormal{surf}$ (blue triangles) can be reproduced with ML+vdW$^\textnormal{surf}$ (red crosses). The deviations are in the range of $\pm$0.1{\AA} vertical adsorption height. Considering that none of the molecules were featured in the training dataset, this demonstrates the increased transferability that the model inherits due to the separate treatment of long- and short-range interactions.
The molecules that lead to the largest deviations in adsorption heights are azulene and anthracene. 
Besides low computational costs, a further advantage of the proposed method is that the vdW correction can be changed. To demonstrate the flexibility of our method we further relax structures at ML+MBD level and compute the related adsorption heights (dark-red star-like shapes). As can be seen from Fig.\ref{fig:x2o_transferability}b, the adsorption heights are very close to ML+vdW$^{\textnormal{surf}}$. Larger deviations are only seen when it comes to benzene. However, the prediction of ML+MBD is in line with the adsorption height of 2.97 \AA reported in refs. \citenum{benzene_e_h,PhysRevLett.115.036104}. 
%It was previously found that DFT+MBD reports lower adsorption energies and higher adsorption heights than DFT+$^{\textnormal{surf}}$ for aromatic metal-adsorbed systems, \cite{PhysRevLett.115.036104,Maurer2015JCP} which is in line with what we find for ML+vdW$^{\textnormal{surf}}$ and ML+MBD.

\begin{table*}[ht]
    \centering
    \begin{tabular}{l|cc|cc}
         & \multicolumn{4}{c}{Adsorption Energy [eV]}\\ Molecule (Symmetry) 
         & DFT+vdW$^{\text{surf}}$ & ML+vdW$^{\text{surf}}$&DFT+MBD &ML+MBD \\ %&Exp.\\%&optB88-vdW\\
          \hline \hline 
         Benzene (hcp0)~\cite{benzenesymm,benzene_e_h,PhysRevLett.115.036104} &-0.75& -0.81 &-0.57&-0.77 \\ %&-0.68 \\ %benzene only pbe +0.8 eV 
         Naphthalene (top30)~\cite{naphthalene_azulene_ener}   &-1.08& -1.19 &-0.77&-1.10 \\ %&-1.01\\
         Anthracene (hcp0)~\cite{anthracene}  &-1.38 &-1.53 &-0.93& -1.12\\ % &-\\ %PBE only about +1 eV% no values available according to ref. %\cite{anthracene}
         %Tetracene (top0)~\cite{tetracene} & &-1.93 &&-1.84& 1.47\\% no comparable dft values
         Pentacene (bridge60)~\cite{anthracene,pentacene_ener}  &-2.40 &-2.12&-1.65&-1.79\\% &-2.14\\ %0.5 PBE alone
         Azulene   (top30)~\cite{naphthalene_azulene_ener}  &
        -1.37&-1.22&-0.91&-1.07\\&%&-1.15\\
    \end{tabular}
    \caption{Adsorption energies for benzene, naphthalene, anthracene, pentacene, and azulene, on Ag(111) on the most stable symmetry side based on literature, where negative values correspond to an exothermic process. Literature values are based on PBE+vdW$^{\textnormal{surf}}$.\cite{benzene_e_h,naphthalene_azulene_ener,anthracene,pentacene_ener} %except for tetracene, whereas the adsorption energy is based on optB88-vdW.~\cite{tetracene}.
    Values are compared to those of ML+vdW$^{\textnormal{surf}}$ and ML+MBD using the relaxed structures obtained with the respective method.}
    \label{tab:X2O_extrapolation}
\end{table*}
In addition to adsorption heights, we sought to investigate whether the ML+vdW$^{\textnormal{surf}}$ method can be used to approximate adsorption energies. Table \ref{tab:X2O_extrapolation} shows the computed adsorption energies with both, ML+vdW$^{\textnormal{surf}}$ and ML+MBD. The trends observed in members of the acene family, i.e., increasing adsorption energy with increasing molecular size, can be reproduced with both methods. However, some energies are overestimated, while others are underestimated with respect to DFT+vdW$^{\textnormal{surf}}$, which correlates with adsorption heights being over- and underestimated, respectively, for all structures except for anthracene.
Nevertheless, given the fact that these systems were never seen by the ML models and the small amount of data used to train ML models, the results are encouraging to develop fully transferable ML models for a wide range of physisorbed structures with only little amount of additional data. This could be applied to large-scale screening studies of organic molecules on surfaces and to perform structural pre-relaxations.

\section{Conclusion} 
We have developed an approach for the efficient prediction of long-range-corrected potential energy surfaces and forces based on machine learning (ML) potentials and external long range dispersion corrections based on Hirshfeld atoms-in-molecules partitioning. Different types of long-range van-der-Waals interactions are implemented including the Tkatchenko-Scheffler vdW and MBD methods to describe nanoclusters on surfaces and organic molecules on metal surfaces. 
One of the powerful features is thus that the type of long-range correction can easily be changed, such that different methods can be employed without the need for retraining.

To apply the method for structure pre-relaxations with ML models trained on little data, we additionally incorporated dynamic stopping criteria that take the variance of machine learning predictions into account and ensure the structure relaxation does not run into unreliable territory. 
The method was tested for fast (pre-)relaxations of complex hybrid systems. Firstly, we demonstrated our framework on gold nanoclusters on a diamond (110) surface and showed that by adaptively optimizing the ML models, global structure searches can be enabled that would be computationally too expensive without the use of ML. 

Secondly, we reused data from Ref.~\citenum{Jeindl2021ACSN} of three organic molecules (X2O) on Ag(111) surfaces. The goal of this study was to assess the applicability of ML models based purely on reused data from open data repositories without generating a tailor-made training data set. This reflects the realistic application scenario in which a small set of initial geometry optimizations can be used to construct an ML+vdW model that can computationally expedite structural pre-relaxation. The conducted tests showed not only the power of open data for developing new methods, but also demonstrated that the method can be used to semi-quantitatively predict adsorption heights and energies and to pre-relax challenging starting systems. Finally, we tested the transferability of our model to unseen organic molecules on Ag(111). 

The approach we present is of general utility for the computational surface science community and has the potential to drastically reduce the computational effort of some of the most common tasks in this field. Our data provides evidence that the construction of a more general and transferable structure relaxation model of hybrid organic-metallic interfaces is feasible and potentially desirable, although small (and rough) system-specific models may be more advantageous in many cases.

\section*{Conflicts of interest}
There is no conflict of interest to declare.

%\end{acknowledgement}

\section{Data availability}
Input and output files for all Au@C calculations, comprising the training dataset and the adaptive run calculations, have been uploaded as a dataset to the NOMAD electronic structure data repository and are freely available under DOI: 10.17172/NOMAD/2021.10.28-1~\cite{AuC}. The molecular geometries and corresponding properties of gold nanoclusters on diamond surfaces are saved in a database format provided by the Atomic Simulation Environment~\cite{Larsen2017IOPP}. The data for X2O are obtained from NOMAD.~\cite{A2O,B2O,P2O}
In addition, files to reproduce figures, test data, and additional code to run ML models is available from figshare (10.6084/m9.figshare.19134602). 

\section{Code availability}
All code developed in this work is made available on figshare (https://figshare.com/s/78b54de875cfb9cadbdd) and GitHub including test examples under URL: https://github.com/maurergroup/SchNet-vdW.
The script to generate the Lindh Hessian for geometry initialization is available via FHI-aims.\cite{aims} A few other versions of the Lindh Hessian script are available via the gensec package~\cite{Maksimovgensec} on GitHub: https://github.com/sabia-group/gensec.

\section*{Acknowledgements}

This work was funded by the Austrian Science Fund (FWF) [J 4522-N and Y1157-N36], the EPSRC Centre for Doctoral Training in Diamond Science and Technology [EP/L015315/1] and the UKRI Future Leaders Fellowship programme [MR/S016023/1]. We are grateful for use of the computing resources from the Scientific Computing Research Technology Platform of the University of Warwick (including access to Avon, Orac and Tinis), the EPSRC-funded Northern Ireland High Performance Computing service [EP/T022175/1] for access to Kelvin2, and the EPSRC-funded High End Computing Materials Chemistry Consortium [EP/R029431/1] for access to the ARCHER2 UK National Supercomputing Service (https://www.archer2.ac.uk).

%%%REFERENCES%%%
%\bibliography{refs.bib} %You need to replace "rsc" on this line with the name of your .bib file
%\bibliographystyle{rsc} %the RSC's .bst file
%aipnum4-2.bst 2019-01-14 (MD) hand-edited version of apsrev4-1.bst

\end{document}

\thispagestyle{plain}
\maketitle
\tableofcontents
\newpage 

\section{Machine Learning (ML) Models and Datasets}
For fitting energies, forces, and Hirshfeld volume ratios, SchNet,~\cite{Schutt2018,Schutt2017_double,Schuett2019JCTC} was used and adapted, which is a continuous-filter convolutional neural network.
\subsection{Datasets}
\paragraph{X2O@Ag}
%The training set for X2O@Ag consisted of 8201 data points taken from Ref.~\cite{Jeindl2021ACSN} for training.
%The training set comprised 399 data points of the clean substrate, 1397 data points of P2O (6,13-pentacenequinone) molecules, 2249 data points of A2O (9,10-anthraquinone) molecules, and 4156 data points of B2O (1,4-benzoquinone) molecules. The molecules were either in the gas phase or on Ag(111) surfaces. Each data point contains a super cell with one or more molecules on a surface. The distribution of the data is shown in Fig.~\ref{fig:data}. 
The training set for X2O@Ag consisted of 8,201 data points taken from Ref.~\cite{Jeindl2021ACSN} for training.
Data points for X2O@Ag were collated from 6,773 single point calculations and 208 geometry optimizations. In addition, we had 6 structure relaxations of the different systems as an additional hold-out test set and further 16 structure relaxations of B2O with systems far away from the surface to test the implementation and accuracy of our method. Geometry optimizations of the hold out test set required about twice as many steps as the geometry optimizations in the training set. %in the training took 8 steps on average. The longest geometry optimization required 35 relaxation steps. In contrast, the structure relaxations of the hold-out test set of B2O required 
The training set was split into 6,800 data points for training, 700 data points for validation, and the rest was used for testing. The model hyper-parameters were sampled on a random grid and optimized according to the performance on the validation set. The final model error was reported on the hold-out test set and is summarized in Table~\ref{tab:mae}.

%\begin{figure}[ht]
%    \centering
%    \includegraphics[scale=0.5]{Figures/Fig.S1.png}
%    \caption{Distribution of data points in the X2O@Ag training set taken from Ref.~\cite{Jeindl2021ACSN}. Each data point comprises one or more molecules in the gas phase, on a Ag(111) surface, or the bulk surface without an organic molecule present.}
%    \label{fig:data}
%\end{figure}

\paragraph{Au@C}
%For Au@C models, we carried out 10-20 geometry relaxations of Au clusters of size 15, 20, 30, 35, 40, 45, and 50 atoms on a diamond surface. In total, 62 optimizations were collated to generate the training set. These computations led to an initial training set of 5368 data pints. 4500 data points were used for training, 500 for validation, and the rest for testing. 
As mentioned in the main text, data for Au@C models were obtained from geometry relaxations of Au nanoclusters on a diamond (110) surface. We started with 62 optimizations of cluster sizes of $n=$ 15,20,30,35,40,45, and 50 which led to a total number of 5,368 data points. Of these data points, we used 4,500 data points for training, 500 for validation, and the rest for testing. In addition, 4 geometry optimizations with a cluster sizes of 20, 30, 40, and 50 were kept as a hold-out test set to test the model performance for optimizations.

For refinement of the training set, we carried out global structure search with initially trained ML models with basin-hopping.~\cite{Wales1997JPCA,Wales1999S}
As starting points for basin-hopping with the initial MLIPs, $ML_{init.}$, we have selected Au nanoclusters (NCs) of different sizes, i.e., the sizes that were featured in the training data set ($n=$ 15, 20, 30, 35, 40, 45, 50) and some that were not included ($n=$ 6, 25, 28, 44, 66). A basin hopping run was initiated for each NC size. The initial structures of known NC sizes were randomly selected from the optimized structures generated with DFT, i.e., we used data points that made up the training set. Systems of unknown NC sizes were constructed as before with ASE and were placed on the center of the diamond (110) surface.~\cite{Larsen2017IOPP} This procedure resulted in 231 structure relaxations. Note that one basin-hopping run comprises several structure relaxations. At the end of each optimization, our algorithm prints the model variance, which was used along with the maximum residual force component to assess the reliability of a structures relaxation. Almost all relaxations with unknown cluster sizes resulted in large model variances, i.e., values $\geq$10 eV/{\AA}, which indicate that the models fail for relaxing these systems. Therefore, data points for adaptive sampling were randomly selected from this set of data points. The relaxations of clusters with sizes known to the MLIPs resulted in smaller model variances and maximum residual forces down to 0.05 eV/\AA, hence, those relaxations that resulted in the largest variances were selected to extend the training set. All selected data points were prepared for additional geometry optimizations with DFT. We added each individual step of a geometry optimization to the training set. In total, 8,893 data points were collected with this procedure. 

MLIPs after the first adaptive sampling run (denoted as ML$_\textnormal{adapt.1}$) were trained on 7,700 data points for training and 800 data points for validation. The same procedure as before was applied to extend the training set further, but using the ML$_\textnormal{adapt.1}$ model instead of the ML$_\textnormal{init.}$ model for initial structure relaxation. In addition, we carried out 243 single point calculations of structures with the largest model errors to let the model know where not to go during optimizations. We collected a total amount of 9,757 data points and final ML$_\textnormal{adapt.2}$ models were trained on 8,500 data points for training and 800 data points for validation. 

% We selected 27 structures with large deviations between ML models and carried out additional DFT structure relaxations, which led to additional 3525 data points, summing up to a training set of 8893 data points. Four ML models were retrained using 7700 data points for training, 800 for validation, and the rest for testing. 

%With the initial training set, we trained four ML models on energies and forces and one on Hirshfeld volume ratios. With these models, we carried out structure relaxations using basin hopping for nanocluster of different sizes, i.e., some sizes that were learned and some sizes that were not learned. 27 structures were selected based on the variance of the energies of the differently trained ML models and relaxations were carried out with DFT for these systems. These data points were added to the training set and models were retrained on 7700 data points for training and 800 data points for validation. In total, 8893 data points were collected.

\subsection{Training}
\paragraph{Energy and Forces}
Energies and forces were trained with standard SchNet models. The energies and forces that were used for training were obtained after subtraction of van der Waals (vdW) contributions. All reference calculations were carried out with FHI-aims.~\cite{Blum2009,zhang2013numeric} 
 %In SchNet, the energies are modelled as the sum of atomic contributions:
% \begin{equation}
 %    E^{ML} = \sum_A^{N_A} E_A^{ML}
 %\end{equation}
 
As already mentioned, two different systems were tested: gold NCs on diamond (110) surfaces (Au@C) and X2O systems on Ag(111) surfaces (X2O@Ag). The energies and forces were trained atom-wise and energies of the whole systems were obtained by summing up atomic contributions. As can be seen from equation 3 in the main text, the resulting energies were mapped to the reference energies. As the systems in the training set were very diverse, total energies varied by a few megaelectronvolts between systems. Thus, energies had to be pre-processed in addition as the current version of SchNet uses data sets saved in an Atomic Simulation Environment (ASE) .db format, which only allows single precision. For X2O@Ag systems we trained energies in the following way:
 \begin{equation}
     E_{\textnormal{training}} = E_{\textnormal{total,vdW-free}}-\sum_\textnormal{A}^{N_A} E_\textnormal{A}.
 \end{equation}
$N_\textnormal{A}$ denotes the number of atoms in a system. 
The atomic energies that were used for scaling were obtained from reference calculations with the same method that was used to generate the training set, i.e., DFT+vdW$^{\text{surf}}$ (see section 2.2.2 in the main text).
 
Due to the large size of the Au@C systems, the energy deviations between the systems ranged from a few to about 100 MeV. Different ways were tested to train the vdW-free energies and forces. The best model performance was obtained when subtracting the minimum of each cluster size individually. The respective values were saved in the data base and could be added subsequently for predictions.
The errors on a hold-out test set for each system for energies and forces can be found in Table \ref{tab:mae}.
After the second adaptive sampling run, a smooth $L_1$ loss function was applied for training. This was done as the training set for ML$_\textnormal{adapt.2}$ contained data points with comparably larger forces and energies than most of the data points. Using the $L_2$ loss function for this dataset would mean that these data points would be weighed comparably large during training, hindering a meaningful model training. Therefore, whenever the model error on a given data point exceeded the mean of the model error on a given batch size, we switched to the $L_1$ loss function. The total loss function for energies and forces for ML$_\textnormal{adapt.2}$ thus reads for a given batch size:
\begin{equation}
L^{\textnormal{batch}} = \begin{cases}L_2 ~~\text{if} ~~ \max\left(\left|E_{\textnormal{local}}^{\textnormal{QC}}-E_{\textnormal{local}}^{\textnormal{ML}}\right|\right) < 3\times \textnormal{mean}\left(\left|E_{\textnormal{local}}^{\textnormal{QC}}-E_{\textnormal{local}}^{\textnormal{ML}}\right|\right) \\
    L_1 ~~\text{if} ~~ \max\left(\left|E_{\textnormal{local}}^{\textnormal{QC}}-E_{\textnormal{local}}^{\textnormal{ML}}\right|\right) \geq 3\times \textnormal{mean}\left(\left|E_{\textnormal{local}}^{\textnormal{QC}}-E_{\textnormal{local}}^{\textnormal{ML}}\right|\right) \end{cases}
\end{equation}

with
\begin{equation}
    L_2 =t_E\left|\left| E_{\textnormal{local}}^{\textnormal{QC}} - E_{\textnormal{local}}^{\textnormal{ML}} \right|\right|^2 + t_F \left|\left| F_{\textnormal{local}}^{\textnormal{QC}}-\dfrac{\partial E_{\textnormal{local}}^{\textnormal{ML}}}{\partial \mathbf{R}}\right|\right|^2 
\end{equation}

and \begin{equation}
        L_1 =t_E\left| E_{\textnormal{local}}^{\textnormal{QC}} - E_{\textnormal{local}}^{\textnormal{ML}} \right| + t_F \left| F_{\textnormal{local}}^{\textnormal{QC}}-\dfrac{\partial E_{\textnormal{local}}^{\textnormal{ML}}}{\partial \mathbf{R}}\right|.
\end{equation}
$E_{\textnormal{local}}^{\textnormal{QC}}$ and $E_{\textnormal{local}}^{\textnormal{ML}}$ denotes a vector of all energies within a given batch size. Different thresholds between 1-10 were tried for switching between $L_1$ and $L_2$ with no significant differences in training performances, hence the original choice of 3 was retained.

Note that the Au@C models obtained after adaptive sampling runs 2 and 3 includes geometries that are unlikely to be visited, but are included in the training to let the model know where not to go. Thus, the MAE and RMSE are expected to increase, which does not imply that the performance of the models for geometry optimizations and global structure searches deteriorates. In fact, if we remove 8 outliers from the computation of the MAE and RMSE, the MAE and RMSE for the energy of the "Au@C adaptive2" and "Au@C adaptive3" models decreases by about a third (MAE) and a tenth (RMSE), respectively, and the MAE and RMSE of forces up to half (MAE) and a third (RMSE), respectively, making the errors comparable to previous adaptive sampling runs.
%In addition to Fig. 2 in the main text, we show the scatter plots including the beforementioned outliers in Fig. \ref{fig:scattersi}.
%\begin{figure}[ht]
%    \centering
%    \includegraphics{Figures/FigS2.png}
%    \caption{Scatter plots for gold nanoclusters (NCs) on diamond (110) surfaces (Au@C) including 8 highly unfavourable structures for energies (left) and forces (right) in addition to Fig. 2 in the main text. Scatter plots for energies and forces are obtained from four trained machine learning (ML) models ("Model 1", "Model 2", "Model 3", and "Model 4"). Note that due to the large energy range, i.e., several million eV, that is spanned by the Au NCs on diamond (110), the energy values are shifted, such that each energetically lowest lying cluster lies at 0 eV.}
%    \label{fig:scattersi}
%\end{figure}

\paragraph{Hirshfeld Volume Ratios}
The Hirshfeld volume ratios were obtained by dividing the effective atom-in-molecule volumes with the free atomic volumes as given in the main text in equations (1) and (2). Hirshfeld volume ratios were trained atom-wise in a single SchNet model. The SchNet output layer was adapted to fit Hirshfeld volume ratios per atom in one neural network, i.e., in a multi-state neural network, by removing the last pooling layer. The last pooling layer usually sums or averages over the atomic contributions, which is not needed in this case.
Hence, multiple, atom-wise values entered the loss function and were mapped directly to the Hirshfeld volume ratios instead of the sum or average of these values.
The errors on a hold-out test set for each system are reported in Table \ref{tab:mae}. 

\paragraph{Model Parameters: X2O@Ag}

For learning energies and forces, a cutoff of 6\,{\AA} was used to represent the atoms in their chemical and structural environments. Larger cutoffs were tested, but did not lead to better results, which was expected as long-range interactions were excluded from the training data. We used default parameters in most cases, hence we only state the model parameters that differed from the default: 128 features, 4 SchNet interaction layers to learn the representation, a learning rate of 3$\times 10^{-4}$, and a batch-size of 8 was used. In total, we trained 4 similar models on energies and forces that differed in the trade-off, used to weight energies ($t$) and forces ($1-t$) during training. Energies were weighted with factors 0.01, 0.03, 0.03, and 0.05 for the different models and the respective force weights were 0.99, 0.97, 0.97, and 0.95.

For learning Hirshfeld volume ratios, a cutoff of 8\,{\AA}, a batch size of 2, and a learning rate of $2\times 10^{-4}$ was used. %For training Hirshfeld volume ratios of the X2O training set, 11 data points were sorted out that turned out to be problematic for training Hirshfeld volume ratios. These data points represented systems with over 400 atoms, while the rest of the structures contained less than 400 atoms per unit cell.
\paragraph{Model Parameters: Au@C}

For training energies and forces, we used a batch size of 4, 4 interaction layers and 128 features to learn the SchNet representation. A learning rate of $2\cdot 10^{-4}$ was used and the weights for the energies were set to 0.03, 0.04, 0.04, and 0.05 with weights for forces set to 0.97, 0.96, 0.96, and 0.95, respectively. Besides, we used default parameters of SchNet.

For training Hirshfeld volume ratios, a cutoff of 6\,{\AA}, a batch size of 4, a learning rate of $5\cdot 10^{-4}$, 4 interaction layers to fit the SchNet representation, 128 features, and 25 Gaussian functions for the input layer were used. The rest of the parameters were set to the default values of SchNet.

\subsection{Model Validation}
The accuracy of the models for X2O@Ag and Au@C are given in Table \ref{tab:mae}. In total, 4 energy and force models and one Hirshfeld model were trained for each data set. The errors are reported on a hold-out test set.
\begin{table}[ht]
    \centering
    \begin{tabular}{c|c|c|c}
        \textbf{System} & Energy [eV] & Forces [eV/{\AA}]& $H_\textnormal{A}$ \\
          \textbf{X2O@Ag}& MAE (RMSE)&MAE (RMSE) & MAE (RMSE)\\
         \hline \hline
         Model 1 & 0.017 (0.025) & 0.021 (0.035) & \\
         Model 2 & 0.018 (0.026) & 0.025 (0.041) & \\
         Model 3 & 0.021 (0.030) & 0.025 (0.041) & \\
         Model 4 & 0.021 (0.030) & 0.024 (0.041) & \\
         $H_\textnormal{A}$ Model & & & 1.4$\cdot10^{-4}$ (7.3$\cdot10^{-4}$) \\ \hline
         
    \textbf{Au@C Initial}\\\hline\hline  
    Model 1 & 0.013 (0.030) & 0.021 (0.046) & \\
       Model 2 & 0.011 (0.029) & 0.020 (0.062) & \\
       Model 3 & 0.011 (0.026) & 0.015 (0.037) & \\
         Model 4 & 0.013 (0.031) & 0.019 (0.050) & \\

         $H_\textnormal{A}$ Model &&&8.1$\cdot10^{-5}$ (1.7$\cdot10^{-4}$)\\\hline
\textbf{Au@C adaptive1} \\ \hline \hline
    Model 1 & 0.020 (0.070) & 0.014 (0.032) & \\
    Model 2 & 0.023 (0.090) & 0.014 (0.033) & \\
    Model 3 & 0.037 (0.055) & 0.021 (0.042) & \\
    Model 4 & 0.028 (0.058) & 0.021 (0.041) & \\ 
    $H_\textnormal{A}$ Model &&&$3.9\cdot10^{-5}$ ($1.3\cdot10^{-4}$)\\\hline
\textbf{Au@C adaptive2} \\ \hline \hline
    Model 1 & 0.091 (0.561) & 0.031 (0.116) & \\
    Model 2 & 0.138 (0.842) & 0.054 (0.260) & \\
    Model 3 & 0.175 (0.998) & 0.062 (0.155) & \\
    Model 4 & 0.138 (0.869) &  0.056 (0.254)& \\ 
    $H_\textnormal{A}$ Model &&&6.2$\cdot10^{-5}$ ($1.7\cdot10^{-4}$)\\
    \textbf{Au@C adaptive3} \\ \hline \hline
    Model 1 & 0.092 (0.55) & 0.040 (0.16) & \\
    Model 2 & 0.094 (0.55) & 0.050 (0.13) & \\
    Model 3 & 0.12 (1.1) & 0.035 (0.26) & \\
    Model 4 & 0.12 (0.59) &  0.066 (0.21)& \\ 
    $H_\textnormal{A}$ Model &&&1.1$\cdot10^{-4}$ ($2.2\cdot10^
    {-4}$)\\
    \end{tabular}
    \caption{Mean absolute errors (MAEs) and root mean-squared errors (RMSEs) of energies, forces, and Hirshfeld-volume ratios on a hold-out test set for X2O@Ag and Au@C. }
    \label{tab:mae}
\end{table}

\clearpage
%\section{ASE Implementation}

%\section{Adaptive sampling for Au@C}
\clearpage
\section{ML Optimization}
%\subsection{Stopping criteria}
The ML models were used for pre-relaxations in case of X2O@Ag and adaptive sampling was carried out for Au@C with initially trained ML models. Thus, as mentioned in the main text briefly, the usually applied \textit{fmax} value of 0.05\,eV/{\AA} could not be reached reliably in all structure relaxations, especially when global structure search was used for adaptive sampling with initial ML models.
\begin{figure}[ht]
    \centering
    \includegraphics[scale=1.5]{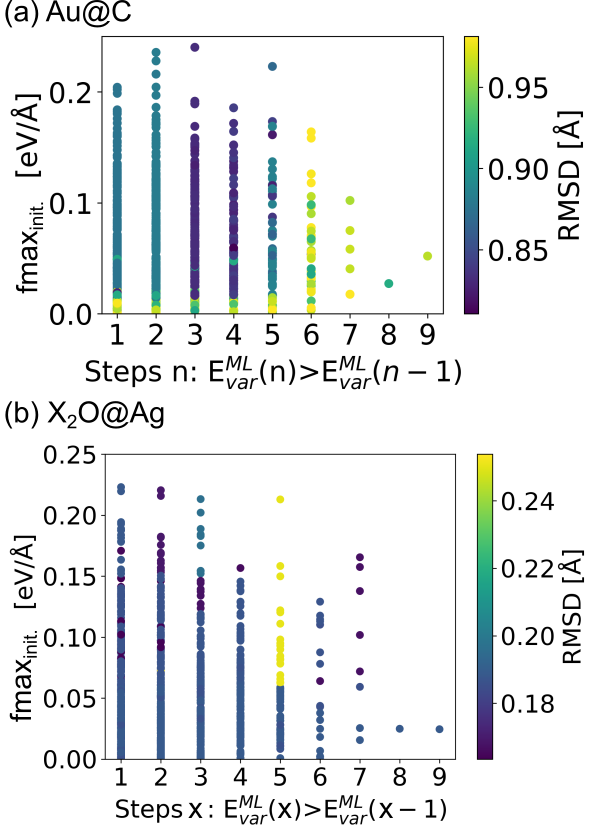}
    \caption{Random grid search of different parameters to stop the structure relaxations with ML models. An initial \textit{fmax}, $fmax_\textnormal{init.}$, and the number of consecutive steps, $x$, after which the variance in energies predicted by the different ML models, $E_{\textnormal{var}}^{\textnormal{ML}}(q)$, was rising, was considered. The color bar shows the root mean squared deviation (RMSD) in {\AA} of the final ML-optimized structure with respect to the DFT-optimized structure. }
    \label{fig:random}
\end{figure}

To this aim we sought to adapt the stopping criteria for structure relaxations to account for the model accuracy. We explored a random grid of 1,000 different stopping criteria %6 additional test set optimizations
using additional structure relaxations of NCs of different sizes for Au@C and the test set of X2O@Ag. We introduced an initial $fmax_{init.}$ in addition to the final \textit{fmax} of 0.05\,eV/{\AA}. Further, we took the number of consecutive steps, $x$, after which the variance in energies, $E_{\textnormal{var}}^{\textnormal{ML}}(q)$, predicted by the query-by-committee models was rising into account. The random grid search is visualized in Fig. \ref{fig:random} (a) and (b) for Au@C and X2O@Ag, respectively. 

As can be seen from Fig.~\ref{fig:random}, in both cases an initial \textit{fmax} in the range of 0.1-0.2\,eV/{\AA} in combination with a preliminary termination of the algorithm after three consecutive steps that showed rising energy variances led to the most stable setup and consequently, to structures that were closest to the DFT minimum (lowest root mean squared deviation (RMSD)). We found that the exact value of the initial \textit{fmax} was not critical, but that it was important to stop the algorithm either after consecutive rising in energy variance or when a final \textit{fmax} of 0.05\,eV/{\AA} was reached. %Noticeably, when the \textit{fmax} was allowed to further decrease, the accuracy of the found structure deteriorated, which is especially prominent for the initial Au@C model.
Independent of the initial $fmax_\textnormal{init.}$, we included another stopping criterion, which terminated the algorithm whenever the model variance exceeded a value of 1\,eV or when the $fmax$ jumped to a value that was larger 2\,eV/{\AA}. Both events were observed when model predictions ran into extrapolative regimes and were not reliable anymore. Note that the model variance rises substantially in extrapolative regions, hence, the threshold of 1\,eV is not critical, but a value of, e.g., 0.5\,eV or 10\,eV would lead to identical results or in the worst case one optimization step fewer or more, respectively.

\clearpage

\subsection{Au@C Optimizations}
\begin{figure}[ht]
    \centering
    \includegraphics[scale=0.9]{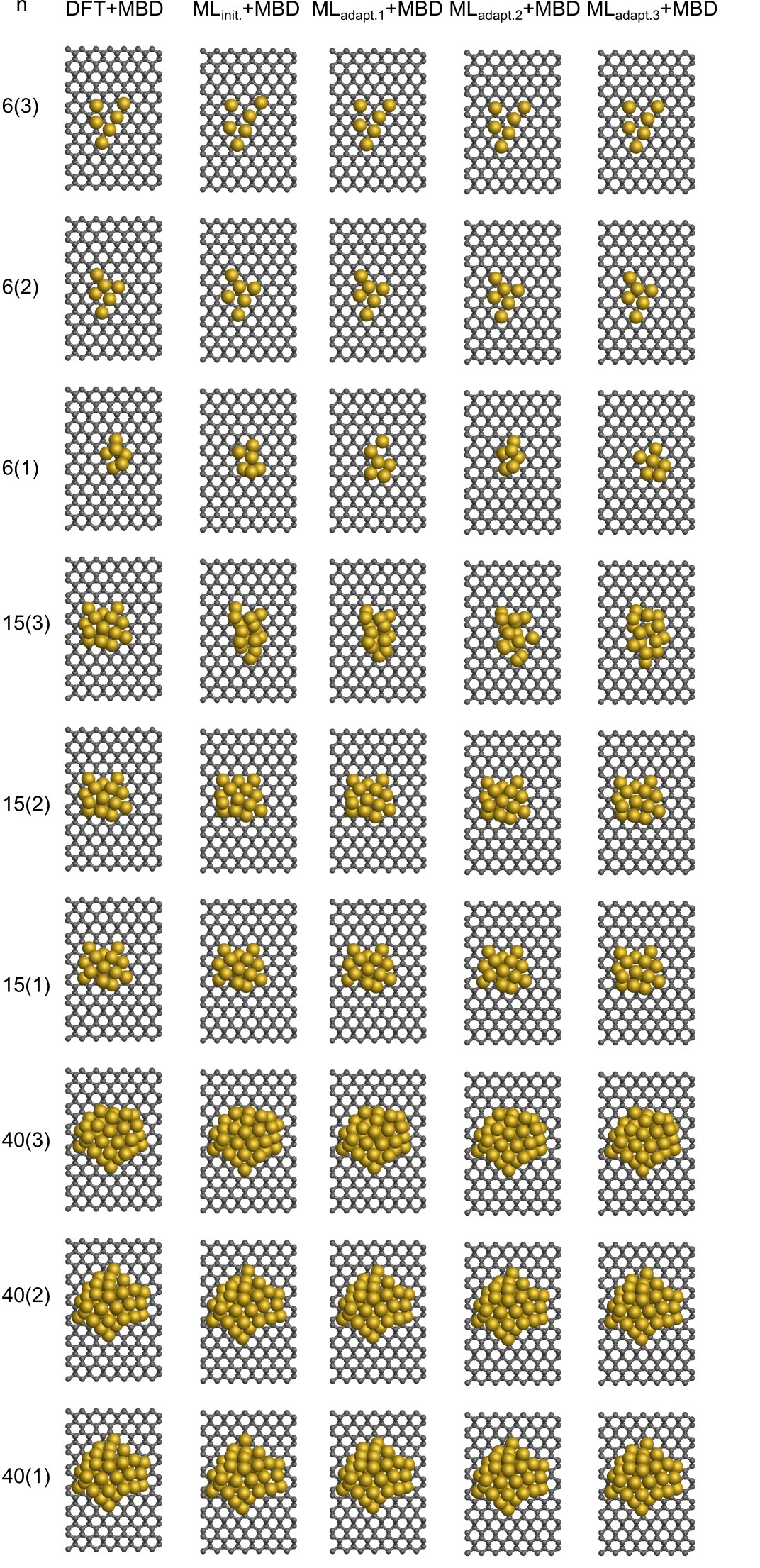}
    \caption{Structures according to Fig. 4d shown from the top view using DFT+MBD, ML$_\textnormal{init.}$+MBD,ML$_\textnormal{adapt.1}$+MBD, and ML$_\textnormal{adapt.2}$+MBD. The number in brackets indicates the energy ranking, i.e., 1 refers to the energetically most favourable structure, while 2 refers to the middle structure and 3 to the energetically least favourable structure.  }
    \label{fig:aucs}
\end{figure}

The structures of the 9 systems with cluster sizes $n=$ 6, 15, and 40 are shown in Fig. \ref{fig:aucs}. The number in brackets indicates the energy ranking, i.e., 1 refers to the energetically most favourable structure, while 2 refers to the middle structure and 3 to the energetically least favourable structure.  

\subsection{X2O@Ag Optimization}
\begin{figure}
    \centering
    \includegraphics[scale=0.6]{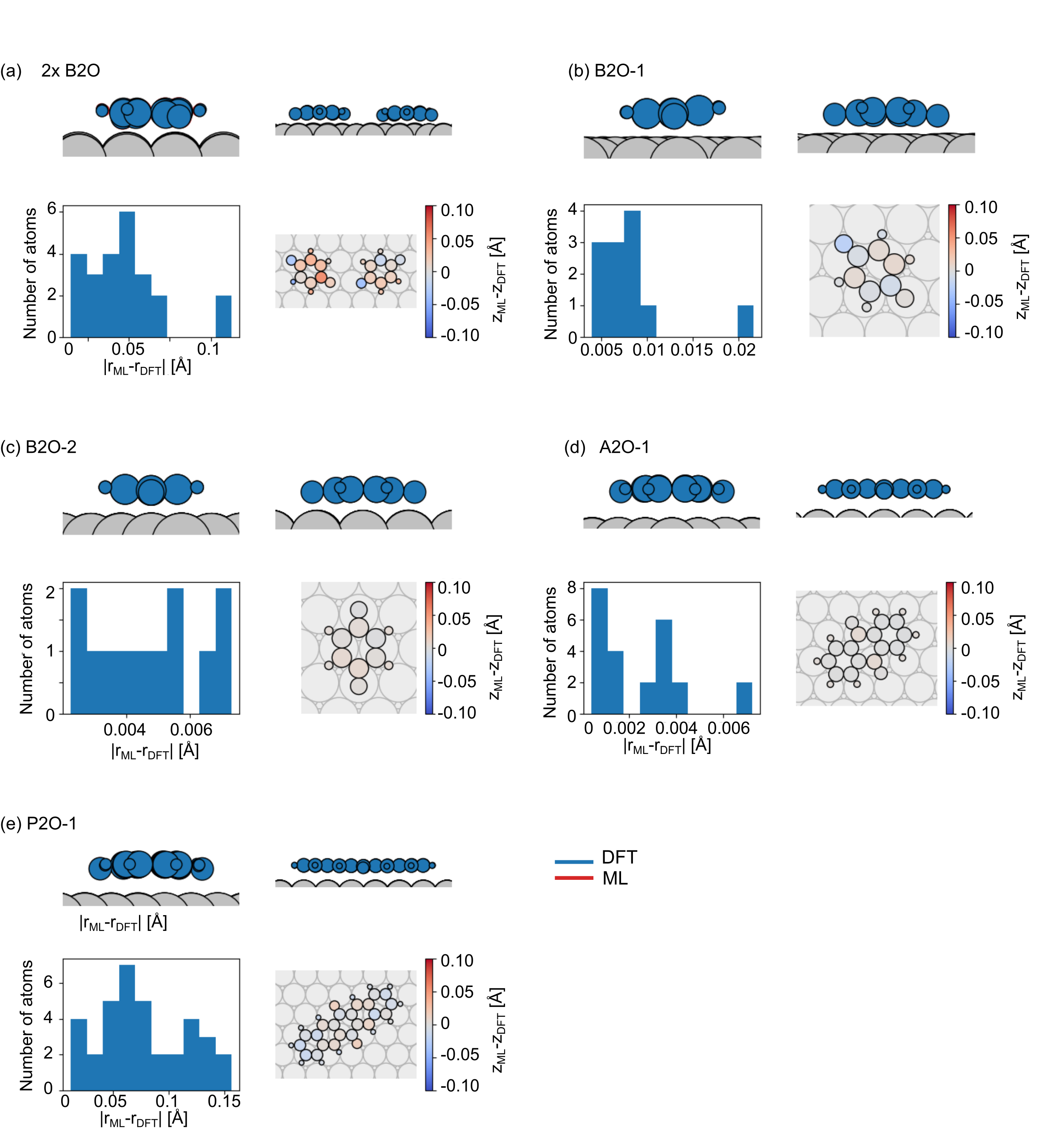}
    \caption{ML+vdW$^{\textnormal{surf}}$ structures compared to DFT+vdW$^{\textnormal{surf}}$ structure for (a) 2x B2O, (b) B2O-1, (c) B2O-2, (d) A2O-1, and (e) P2O-1 according to Fig. 5 in the main text.}
    \label{fig:x2o1}
\end{figure}

The ML-optimized structures of the test set according to Fig. 5 in the main text are assessed in Fig. \ref{fig:x2o1}. 

\begin{figure}[ht]
    \centering
    \includegraphics[scale=0.35]{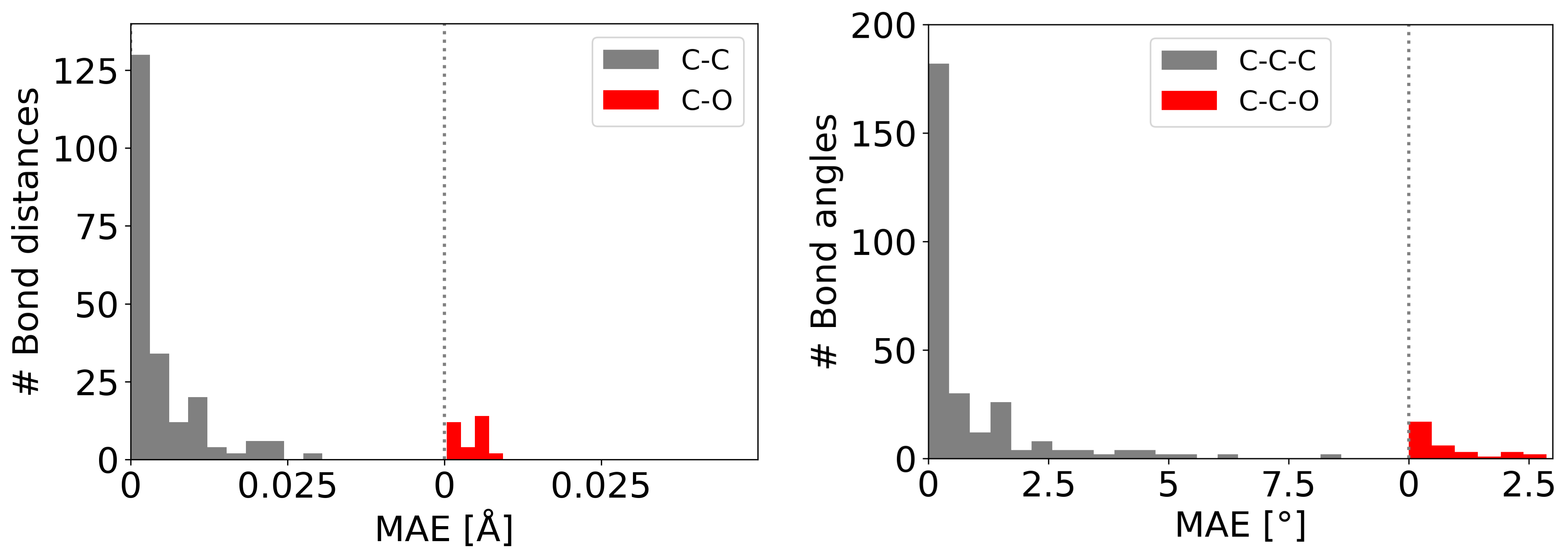}
    \caption{Mean absolute error (MAE) in bond distances (left plot) and bond angles (right plot) of X2O@Ag structures of the test set. }
    \label{fig:x2o_bondangle}
\end{figure}

The errors in bond distances and bond angles of the test set structures is shown in Fig. \ref{fig:x2o_bondangle}.
\clearpage

\providecommand{\latin}[1]{#1}
\makeatletter
\providecommand{\doi}
  {\begingroup\let\do\@makeother\dospecials
  \catcode`\{=1 \catcode`\}=2 \doi@aux}
\providecommand{\doi@aux}[1]{\endgroup\texttt{#1}}
\makeatother
\providecommand*\mcitethebibliography{\thebibliography}
\csname @ifundefined\endcsname{endmcitethebibliography}
  {\let\endmcitethebibliography\endthebibliography}{}

%\bibliographystyle{achemso}
%\bibliography{refs.bib}